%% file: ms.tex
\documentclass{emulateapj}
\def\ah{^{\rm h}}
\def\am{^{\rm m}}
\def\as{^{\rm s}}
\def\pr{^{\prime}}
\def\2pr{^{\prime \prime}}

\def\greatsim{\mathrel{\raise.3ex\hbox{$>$\kern-.75em\lower1ex\hbox{$\sim$}}}}
\def\lesssim{\mathrel{\raise.3ex\hbox{$<$\kern-.75em\lower1ex\hbox{$\sim$}}}}
\def\gs{\mathrel{\raise0.27ex\hbox{$>$}\kern-0.70em % Greater/squiggles
\lower0.71ex\hbox{{$\scriptstyle \sim$}}}}
\def\ls{\mathrel{\raise0.27ex\hbox{$<$}\kern-0.70em % Less than/squiggles
\lower0.71ex\hbox{{$\scriptstyle \sim$}}}}

\slugcomment{Submitted to {\em Astronomical Journal}}

\shorttitle{SNe in High Redshift Galaxy Clusters}
\shortauthors{Dawson et al.}

\input{outlinemacro}

\newboolean{draft}
\setboolean{draft}{false}

\begin{document}

\title{An Intensive
  HST\footnote{Based on observations made with the NASA/ESA Hubble Space
    Telescope and obtained from the data archive at the Space
    Telescope Institute. STScI is operated by the Association of
    Universities for Research in Astronomy, Inc. under the NASA
    contract NAS 5-26555.  The observations are associated with
    program 10496.} $\,$Survey for $z>1$ Type Ia Supernovae by Targeting Galaxy Clusters}

\author{
K.~S.~Dawson\altaffilmark{2,3},
G.~Aldering\altaffilmark{3},
R.~Amanullah\altaffilmark{4},
K.~Barbary\altaffilmark{3,4},
L.~F.~ Barrientos\altaffilmark{5},
M.~Brodwin\altaffilmark{6,7},
N.~Connolly\altaffilmark{8},
A.~Dey\altaffilmark{9},
M.~Doi\altaffilmark{10},
M.~Donahue\altaffilmark{11},
P.~Eisenhardt\altaffilmark{12},
E.~Ellingson\altaffilmark{13},
L.~Faccioli\altaffilmark{3},
V.~Fadeyev\altaffilmark{14},
H.~K.~Fakhouri\altaffilmark{3,4},
A.~S.~Fruchter\altaffilmark{15},
D.~G.~Gilbank\altaffilmark{16},
M.~D.~Gladders\altaffilmark{17},
G.~Goldhaber\altaffilmark{3,4},
A.~H.~Gonzalez\altaffilmark{18},
A.~Goobar\altaffilmark{19,20},
A.~Gude\altaffilmark{4},
T.~Hattori\altaffilmark{21},
H.~Hoekstra\altaffilmark{22,23,24},
X.~Huang\altaffilmark{4},
Y.~Ihara\altaffilmark{10,25},
B.~T.~Jannuzi\altaffilmark{9},
D.~Johnston\altaffilmark{12,26},
K.~Kashikawa\altaffilmark{27},
B.~Koester\altaffilmark{17,28},
K.~Konishi\altaffilmark{29},
M.~Kowalski\altaffilmark{30},
C.~Lidman\altaffilmark{31},
E.~V.~Linder\altaffilmark{3},
L.~Lubin\altaffilmark{32},
J.~Meyers\altaffilmark{3,4},
T.~Morokuma\altaffilmark{25,27},
F.~Munshi\altaffilmark{4},
C.~Mullis\altaffilmark{33},
T.~Oda\altaffilmark{34},
N.~Panagia\altaffilmark{15},
S.~Perlmutter\altaffilmark{3,4},
M.~Postman\altaffilmark{15},
T.~Pritchard\altaffilmark{3,4},
J.~Rhodes\altaffilmark{12,35},
P.~Rosati\altaffilmark{36},
D.~Rubin\altaffilmark{3,4},
D.~J.~Schlegel\altaffilmark{3},
A.~Spadafora\altaffilmark{3},
S.~A.~Stanford\altaffilmark{32,37},
V.~Stanishev\altaffilmark{19,38},
D.~Stern\altaffilmark{12},
M.~Strovink\altaffilmark{3,4},
N.~Suzuki\altaffilmark{3},
N.~Takanashi\altaffilmark{27},
K.~Tokita\altaffilmark{10},
M.~Wagner\altaffilmark{3,4},
L.~Wang\altaffilmark{39},
N.~Yasuda\altaffilmark{29},
H.~K.~C.~Yee\altaffilmark{40}
(The Supernova Cosmology Project)
}

\altaffiltext{2}{Department of Physics and Astronomy, University of Utah, Salt Lake City, UT 84112}
\altaffiltext{3}{E.O. Lawrence Berkeley National Lab, 1 Cyclotron Rd., Berkeley CA, 94720}
\altaffiltext{4}{Department of Physics, University of California Berkeley, Berkeley, CA 94720}
\altaffiltext{5}{Universidad Catolica de Chile}
\altaffiltext{6}{Harvard-Smithsonian Center for Astrophysics, 60 Garden Street, Cambridge,
MA 02138}
\altaffiltext{7}{W. M. Keck Postdoctoral Fellow at the Harvard-Smithsonian Center for
Astrophysics}
\altaffiltext{8}{Hamilton College Department of Physics, Clinton, NY 13323}
\altaffiltext{9}{National Optical Astronomy Observatory, Tucson, AZ 85726-6732}
\altaffiltext{10}{Institute of Astronomy, Graduate School of Science, University of Tokyo 2-21-1 Osawa, Mitaka, Tokyo 181-0015, Japan}
\altaffiltext{11}{Michigan State University, Department of Physics and Astronomy, East
Lansing, MI, 48824-2320}
\altaffiltext{12}{Jet Propulsion Laboratory, California Institute of Technology, Pasadena, CA, 91109}
\altaffiltext{13}{Center for Astrophysics and Space Astronomy, 389 UCB, University of Colorado, Boulder, CO 80309}
\altaffiltext{14}{Santa Cruz Institute for Particle Physics, University of California, Santa Cruz, CA 94064}
\altaffiltext{15}{Space Telescope Science Institute, 3700 San Martin Drive, Baltimore, MD 21218, USA}
\altaffiltext{16}{Department of Physics and Astronomy,
University Of Waterloo, Waterloo, Ontario, Canada N2L 3G1}
\altaffiltext{17}{Department of Astronomy and Astrophysics, University of Chicago, Chicago, IL 60637}
\altaffiltext{18}{Department of Astronomy, University of Florida, Gainesville, FL 32611-2055}
\altaffiltext{19}{Department of Physics, Stockholm University, Albanova University Center, SE-106 91, Stockholm, Sweden}
\altaffiltext{20}{The Oskar Klein Centre for Cosmo Particle Physics, AlbaNova, SE-106 91
Stockholm, Sweden}
\altaffiltext{21}{Subaru Telescope, National Astronomical Observatory of Japan, 650 North A'ohaku Place, Hilo, HI 96720}
\altaffiltext{22}{Department of Physics and Astronomy, University of Victoria, Victoria, BC, V8W 2Y2, Canada}
\altaffiltext{23}{Leiden Observatory, Leiden, Netherlands}
\altaffiltext{24}{Alfred P. Sloan Fellow}
\altaffiltext{25}{JSPS Fellow}
\altaffiltext{25}{Department of Physics and Astronomy, Northwestern University, 2145 Sheridan Road, Evanston, IL 60208}
\altaffiltext{27}{National Astronomical Observatory of Japan, 2-21-1 Osawa, Mitaka, Tokyo,181-8588, Japan}
\altaffiltext{28}{Kavli Institute for Cosmological Physics, The University of
Chicago, Chicago IL 60637, USA}
\altaffiltext{29}{Institute for Cosmic Ray Research, University of Tokyo, 5-1-5, Kashiwanoha, Kashiwa, Chiba, 277-8582, Japan}
\altaffiltext{30}{Institut f\"ur Physik, Humboldt-Universit\"at zu Berlin}
\altaffiltext{31}{European Southern Observatory, Alonso de Cordova 3107, Vitacura, Casilla 19001, Santiago 19, Chile}
\altaffiltext{32}{University of California, Davis, CA 95618}
\altaffiltext{33}{Wachovia Corporation, NC6740, 100 N. Main Street, Winston-Salem, NC 27101}
\altaffiltext{34}{Department of Astronomy, Kyoto University, Sakyo-ku, Kyoto 606-8502, Japan}
\altaffiltext{35}{California Institute of Technology, Pasadena, CA 91125}
\altaffiltext{36}{ESO, Karl-Schwarzschild-Strasse 2, D-85748 Garching, Germany}
\altaffiltext{37}{Institute of Geophysics and Planetary Physics, Lawrence Livermore National Laboratory, Livermore, CA 94550}
\altaffiltext{38}{CENTRA - Centro Multidisciplinar de Astrof\'isica, Instituto Superior
T\'ecnico, Av. Rovisco Pais 1, 1049-001 Lisbon, Portugal}
\altaffiltext{39}{Department of Physics, Texas A \& M University, College
Station, TX 77843, USA}
\altaffiltext{40}{Department of Astronomy and Astrophysics, University of Toronto, Toronto, ON M5S 3H4, Canada}

\email{kdawson@physics.utah.edu}

\begin{abstract}
We present a new survey strategy to discover and study high redshift
Type Ia supernovae (SNe Ia) using the Hubble Space Telescope (HST). 
By targeting massive galaxy clusters at $0.9<z<1.5$, we obtain a
twofold improvement in the efficiency of finding SNe compared to an HST
field survey and a factor of three improvement in the total yield of SN
detections in relatively dust-free red-sequence galaxies.  
In total, sixteen SNe were discovered at $z > 0.95$, nine of which were
in galaxy clusters.
This strategy provides a
SN sample that can be used to decouple the effects of host galaxy
extinction and intrinsic color in high redshift SNe, thereby reducing
one of the largest systematic uncertainties in SN cosmology.
\end{abstract}

\section{Introduction}\label{sec:intro}
\input{section1}

\section{Supernova Colors and Dust Extinction}\label{sec:dust}
\input{section2}

\section{The HST Cluster SN Survey}\label{sec:search}
\input{section3}

\section{First Results from the HST Cluster SN Survey}\label{sec:results}
\input{section4}

\section{Discussion: SN Production in Clusters}\label{sec:clusters}
\input{section5}

\section{Conclusion}\label{sec:conclusion}
\input{section6}

\bibliographystyle{apj}
\bibliography{archive}

\begin{appendix}

\section{Computing the Red-Sequence Richness}\label{sec:appendix}
\input{appendix}

\end{appendix}

\end{document}

%% file: outlinemacro.tex
\usepackage{ifthen,color,graphicx}

\newcommand{\outlinestart}[1]
{
\ifthenelse{\boolean{#1}}{\begin{enumerate}}{}
}

\newcommand{\outline}[2]
{
\ifthenelse{\boolean{#1}}{\it \color {red} \item #2 \rm \color {black}}{}
}

\newcommand{\outlineend}[1]
{
\ifthenelse{\boolean{#1}}{\end{enumerate}}{}
}

%% file: section1.tex
Type Ia supernova (SN) searches of the 1990s provided the first
hints \citep{perlmutter98a, garnavich98a, schmidt98a} and then the
first strong evidence for cosmological acceleration \citep{riess98a,perlmutter99a}
\citep[for a review see][]{perlmutter03a}.
Analyses of subsequent data sets have steadily improved the
constraints on cosmology.  The analysis of the ``Union'' compilation of
all current SNe Ia combined with the results from the Wilkinson
Microwave Anisotropy Probe (WMAP) \citep{dunkley09a} and baryon acoustic
oscillations (BAO) \citep{eisenstein05a} lead to an estimate of the
density of dark energy $\Omega_{de} = \Omega_\Lambda=
0.713^{+0.027}_{-0.029} (stat)^{+0.036}_{-0.039} (sys)$ in a flat
universe, and a pressure to density ratio
$w=-0.969^{+0.059}_{-0.063}(stat)^{+0.063}_{-0.066} (sys)$ assuming an
equation of state that does not vary with time \citep{kowalski08a}.
From these results it is apparent that adding more SNe will not
significantly improve the constraints on dark energy without first reducing the
systematic uncertainties associated with the SN observations.
Hence, better measured and understood subsamples of SNe are needed to
reduce the systematic uncertainties associated with our understanding
of dark energy.

SN Ia observations remain the most accurate technique to measure the
expansion history of the universe, with dedicated surveys covering the
full redshift range out to $z \sim 1.5$.  The Katzman Automatic
Imaging Telescope (KAIT) \citep{filippenko01a}, the Nearby SN
Factory \citep{aldering02a}, the Center for Astrophysics SN group (CfA3 sample)
\citep{hicken09a} and the Carnegie SN
Project \citep{hamuy06a} are conducting ground-based surveys of SNe
at low redshifts ($z<0.1$).  Programs such as the SDSS SN Survey ($0.1
< z < 0.3$) \citep{frieman07a}, the SN Legacy Survey
(SNLS) \citep{astier06a} and ESSENCE \citep{wood-vasey07a} ($0.3 < z <
0.8$) are building samples of several hundred SNe at intermediate
redshifts using photometry from two to four meter class telescopes.
The ground-based discovery of SN~1998eq \citep{aldering98a} with the
Low Resolution Imaging Spectrometer \citep[LRIS;][]{oke95a} on Keck II
was the first SN discovered at a redshift $z>1$.
The discovery was soon followed by ground-based discoveries of the $z>1$ SNe
SN~1999fk, SN~1999fo, SN~1999fu, and
SN~1999fv \citep{coil00a,tonry99a}.  SN searches at the highest redshifts
are now done primarily with the Hubble Space Telescope (HST), where
high resolution and low background provide the most precise
measurements of distant SNe.  However, the total number of
well-observed high redshift SNe remains small, with only 23 SNe
lightcurves from an HST survey of the GOODS\footnote{Based
on observations made with the NASA/ESA Hubble Space Telescope,
obtained from the data archive at the Space Telescope Institute. STScI
is operated by the Association of Universities for Research in
Astronomy, Inc. under the NASA contract NAS 5-26555.  The observations
are associated with program 9425.}
fields contributing to the full sample at
$z>1$ \citep{riess07a}.  In addition, at $z>1$, even measurements with
HST are limited by both the statistical and systematic uncertainties
in applying extinction corrections due to dust in galaxies.

Discovering SNe~Ia in relatively dust-free early-type galaxies offers
an opportunity to reduce the uncertainty related to extinction corrections.
SNe in these galaxies can therefore reduce the systematic uncertainty in estimating
cosmological parameters.
The centers of rich galaxy clusters are dominated by
such galaxies, and galaxy clusters up to $z\sim1.5$ are known.
SNe in this redshift range probe the gradual transition to 
the decelerating, matter-dominated high redshift universe.
Galaxy clusters also provide a significant enhancement in the density
of potential SN hosts in a relatively small field of view.
For this reason, cluster fields were targeted in early SN
surveys \citep[e.g.][]{perlmutter95a,reiss98a}.
Although this strategy became less advantageous with wide field capabilities
on ground-based telescopes, the expected boost in the yield of SNe due to the presence of
galaxy clusters is still significant in a survey with HST.
Discovering SNe~Ia in the early-type galaxies of very distant
galaxy clusters with HST is an approach that we develop in this paper.

In a collaboration with members of the IRAC Shallow Cluster
Survey \citep{eisenhardt08a}, the Red-Sequence Cluster Survey (RCS)
and RCS-2 \citep{gladders05a,yee07a},
the XMM Cluster Survey \citep{sahlen08}, the
Palomar Distant Cluster Survey \citep{postman96a}, the XMM-Newton
Distant Cluster Project \citep{bohringer05a}, and the ROSAT Deep Cluster Survey
(RDCS) \citep{rosati99a}, the Supernova Cosmology
Project (SCP) developed and carried out a novel survey approach to
improve the efficiency and usefulness of high redshift SN observations
with HST.
We used the Advanced Camera for Surveys (ACS) to search for and
observe SNe in recently discovered massive galaxy clusters
in the redshift range $0.9 < z < 1.5$.
Data from this program will be used for cosmological constraints,
for a comparison study of
host galaxy environment with respect to SN properties, to estimate SN
rates in clusters at high redshift and for cluster studies
with results that will appear in separate publications.
In this paper we present this
new survey strategy and report the results of its first use, a program
called the HST Cluster SN Survey (Program number 10496, PI: Perlmutter).

In \S\ref{sec:dust} we discuss the survey approach
in the context of SN color and extinction.  We describe the
observations and search for SNe in \S\ref{sec:search}.  The SN yield
from this program is reported in \S\ref{sec:results} and an analysis
of the expectations of increased yield from the distribution and
richness of red-sequence cluster galaxies is presented
in \S\ref{sec:clusters}.  Finally, a summary and implications for
future work are presented in \S\ref{sec:conclusion}.

%% file: section2.tex
Empirical studies have shown that the absolute luminosity of
a SN~Ia is correlated with the shape of the SN lightcurve (e.g.,
$\Delta m_{15}$ in \citet{phillips93a}, $\Delta$ in \citet{riess95a,
riess96b, jha07a}, and the stretch parameter $s$ in \citet{perlmutter95a,
perlmutter97a}) and the SN color \citep{tripp98a, phillips99a,
guy05a}.  Fainter SNe~Ia tend to have redder colors at maximum
light and/or narrower lightcurves. 
These correlations are used to improve SNe~Ia distances and to reduce
the dispersion in the Hubble diagram.

While it is clear that extinction by host galaxy dust does redden and
dim some SNe~Ia, there is no consensus on how to differentiate
reddening by dust from an intrinsic relation between SN color and
luminosity.  Most studies \citep[e.g.][]{tripp98a,astier06a,nobili08a,conley07a}
have shown that the observed relation between SN color and luminosity is considerably
different from the average relation observed for dust reddened objects
in our galaxy. This could mean that the dust
obscuring SNe~Ia in external galaxies is different from the dust in our
galaxy (which would make our galaxy rather special).
It may also mean that the
intrinsic relation between SN color and luminosity dominates
the observed trend. A sphere of dust surrounding
the explosion site is a third possibility proposed by \citet{goobar08a}.

The systematic errors that are associated with the interpretation of
the color luminosity relation are now comparable to the statistical errors. 
For example, changing $R_V$ by one
(approximately the difference between the value of $\beta$ used in the SALT
analysis of SNLS and the value of $R_B$ measured in our own galaxy)
changes w by 0.04 when considering the typical range of color with redshift found in
supernova data sets \citep{kowalski08a,wood-vasey07a}.
The systematic uncertainty associated
with any prior used on the distribution of the amount of extinction
is at least as big.
Finally, the current samples of SNe host environments are heterogeneous, making it
difficult to differentiate the intrinsic color-luminosity relationship
from host-galaxy extinction.
All of these sources of systematic uncertainties are similar in size or larger than the statistical
uncertainty in w measured for the SCP Union Compilation \citep{kowalski08a}.

\subsection{A New Survey Strategy}\label{subsec:dustfreegalaxies}

To reduce the systematic errors inherent to the extinction
correction, we designed a new strategy to collect a uniform sample of
SNe in host galaxies anticipated to have negligible extinction from
galactic dust.

The new strategy targets galaxy clusters that provide a rich
environment of elliptical galaxies. Direct observations indicate that
dust is found in about $50\%$ of nearby elliptical galaxies, but the
amount of dust is usually very small and confined to the central few
hundred parsecs
\citep{tran01a}.  Spitzer data \citep{temi05a} confirm that most
nearby ellipticals have the spectral energy distribution (SED) expected for dust-free systems.  The
clearest line of evidence that dust has little effect on the colors of
elliptical galaxies comes from observations of the uniformity of the
red-sequence in galaxy clusters.  A
dispersion in the color-magnitude relation in elliptical cluster
members of roughly $3\%$ is observed in cluster galaxies ranging from
Coma to intermediate redshifts \citep{bower92a,ellis97a,stanford98a}.
This relation has recently been shown to hold for early-type galaxies
in smaller clusters and groups as well \citep{hogg04a}.  
The observed dispersion is attributed to differences in the age of
stellar populations and recent
results show that the same small dispersion in color extends to
redshifts $z>1$ \citep{blakeslee03a,mei06b,mei06a,lidman08a,mei08a}.
Another line of evidence for lack of dust
comes from observations of quasars behind clusters - \citet{chelouche07a} find an average
extinction of less than 0.01 magnitudes when comparing these quasars to quasars in the field.
The small amount of dust and uniform galaxy population implied by the
observations of the red-sequence makes clusters an ideal environment
to obtain direct measurements of SNe.
Moreover, each cluster SN can be tested for dust by comparing the
photometric and morphological properties of the host to the galaxies
that comprise the red-sequence.

It has been shown that the intrinsic dispersion in cosmological
distances derived from SNe depends on host galaxy
type \citep{sullivan03a,jha07a}.  SNe hosted by irregular galaxies
show the highest intrinsic dispersion, while SNe hosted by passive,
early-type galaxies have the lowest.  The increased dispersion in SN
distance estimates has been attributed to increased levels of dust
due to star-formation, as well as a more heterogeneous
progenitor population.

Additional observations indicate that SNe Ia properties 
are correlated with host galaxy properties.
The average SN Ia hosted by a passive galaxy has a
faster rise and fall than the average
SN Ia hosted by a star-forming galaxy \citep{sullivan06a}.  
The rate of SN Ia production per unit mass increases by a factor of $\sim 20$ across the
Hubble sequence from passive to irregular galaxies \citep{mannucci05a}.
Finally, \citet{hamuy00a,takanashi08a} show that a subsample of
elliptical-hosted SNe populate a much smaller area of color-stretch
parameter space than the full sample of SNe.

By targeting massive galaxy clusters and the associated overdensities
of early-type and late-type galaxies, we develop a technique that boosts
the yield of SNe at a specific and controlled redshift.
Using the most distant clusters known at the time of the program,
the HST Cluster SN Survey yields a much higher number of $z>1$ SNe than would be expected
for a blank-field search with the same areal coverage.  In addition,
this survey strategy minimizes the systematic uncertainty resulting from
the degeneracy between intrinsic SN color and extinction from dust.
This program also permits a study of SN 
progenitors in the richest environments in the Universe.
A large sample of elliptical-hosted SNe characterized by
the cluster red-sequences therefore
reduces several of the systematic errors associated with a broader
sample of SNe.

%% file: section3.tex
Massive high redshift galaxy clusters were observed using ACS on HST
from July 2005 until December 2006.  Clusters were chosen from X-ray,
optical, and IR surveys and cover a redshift range $0.9 < z < 1.5$.
Of the 25 clusters in the sample, 23 have spectroscopically confirmed
redshifts.  The remaining two clusters have redshift estimates from the
photometric properties of the galaxies in the overdense region.
Cluster positions, redshifts, and references can be found in
Table \ref{table:clusters}.

\input{table1}

The survey strategy was expected to produce more than one SN per field
and an elliptical-hosted SN Ia in every other field due to the
presence of clusters in addition to field galaxies.  The high rate of
SNe allowed a simple strategy for scheduling search and follow-up
observations.  Clusters around a redshift $z=1$ were observed
approximately once every 26 days.  Clusters at redshifts $z>1.3$ were
observed once every 20 days to increase the number of photon counts
per lightcurve for comparable constraints on these fainter, more distant
SNe.  The average visibility window in two-gyro mode was four to five
months, resulting in an average of 7-8 HST visits per cluster.

Using a reduction pipeline that will be described in detail in
Suzuki et al. (2009, in preparation), four dithered exposures in the F850LP
(hereafter $z_{850}$) filter were used for the pre-scheduled search and follow-up
observations.  Pixels contaminated by cosmic rays were identified and de-weighted
in the individual exposures.  The four exposures were stacked using
MultiDrizzle \citep{fruchter02a,koekemoer02a}.
For the work presented here, an output pixel scale of $0.05\2pr$ was used.
This strategy also included a fifth exposure in the F775W (hereafter
$i_{775}$) filter.  The second band was intended to study the
intrinsic color variation of SNe Ia and to measure the red-sequence in
the galaxy clusters.  Thus by the end of the survey, modulo end
effects, every SN was observed with a fully sampled lightcurve in two
bands even without additional follow-up observations.  A summary of
the observations can be found in Table~\ref{table:obstimes}.  A more
detailed description of the targets is found in \S\ref{sec:clusters}
and Appendix~\ref{sec:appendix}.

\input{table2}

Immediately following each exposure, the $z_{850}$ image was searched
for SNe using the initial visit as a reference following the technique
developed for the earliest SCP
surveys \citep{perlmutter95a,perlmutter97a} and used most recently
in \citet{kuznetsova08a}.  Supernovae at $z>1$ were typically
discovered before maximum brightness, at a Vega magnitude of roughly
$24-24.5$ with a signal-to-noise ratio of $\sim 6-15$ depending on the
discovery phase and the level of host-galaxy contamination.
The observations in this survey were spaced in time to provide at least three lightcurve points
around maximum light in the $i_{775}$ and $z_{850}$ filters.
Observations of a typical SN Ia provide the data for lightcurve
fits with an uncertainty of 0.06 on the rest-frame B-band peak magnitude and 0.05
on the stretch parameter at $z=1$.
A typical SN in this program has an uncertainty of 0.08
on the rest-frame B-band magnitude and 0.08 on the stretch parameter around $z=1.3$.

Upon discovery of a supernova, we obtained ground-based spectroscopy using
pre-scheduled time on Keck or Subaru, or using queue mode observations
at VLT.
Spectroscopy was used primarily for
precise redshifts and confirmation of galaxy type and in some cases
for a spectral typing of the SN itself.  Because the old stellar
populations of early-type galaxies overwhelmingly host
SNe~Ia \citep{cappellaro99a,hakobyan08a}, host spectra were used to
infer SN type.  By targeting early type galaxies, this strategy
therefore avoids the expensive ToO grism observations on ACS that are
typically required to determine the type of high redshift SNe.  Orbits
that would normally be used for spectroscopy were used instead to
search additional fields for SNe.
After discovery and confirmation that a SN is at a redshift $1.0 < z
<1.3$, we scheduled one additional NIC2 F110W observation within
$\pm$5 days of the SN lightcurve peak.  Full NICMOS lightcurves
consisting of between four and nine orbits were obtained for the
highest redshift SNe Ia ($z>1.3$).

In total, the program consisted of 188 pre-scheduled orbits with ACS
and 31 follow-up orbits with NICMOS.  We dedicated 19 nights with the
Faint Object Camera and Spectrograph \citep[FOCAS:][]{kashikawa02a} on
Subaru, four nights with the LRIS on Keck I, four nights with the Deep Imaging
Multi-Object Spectrograph (DEIMOS: \citet{faber03a}) on Keck II, and 32
hours of queue mode with the Focal Reducer and Low Dispersion
Spectrographs (FORS1 and FORS2: \citet{appenzeller98a}) on Antu and
Kueyen (Unit 1 and Unit 2 of the ESO Very Large Telescope(VLT)) for
spectroscopic follow-up observations.

%% file: table1.tex
\begin{deluxetable*}{lcccccc}
\tablewidth{0pt}
%\tabletypesize{\small}
\tabletypesize{\scriptsize}
\tablecaption{\label{table:clusters} Cluster positions and redshifts}
\tablehead{\colhead{ID} & \colhead{Cluster} & \colhead{R.A. (J2000)} & \colhead{Decl. (J2000)} & \colhead{Redshift} & \colhead{Discovery} & \colhead{$N_{red}$\tablenotemark{b}}}
\startdata
A & XMMXCS J2215.9-1738 & $22\ah\,15\am\,58.5\as$ & $-17^{\circ}\,38\pr\,02\2pr$ & 1.45 & X-ray &  $ 17.3 \pm 3.1$\\
B & XMMU J2205.8-0159 & $22\ah\,05\am\,50.7\as$ & $-01^{\circ}\,59\pr\,30\2pr$ & 1.12   & X-ray &  $  4.9 \pm 2.2$\\
C & XMMU J1229.4+0151 & $12\ah\,29\am\,28.8\as$ & $+01^{\circ}\,51\pr\,34\2pr$ & 0.98   & X-ray &  $ 35.9 \pm 2.6$\\
D & RCS022144-0321.7  & $02\ah\,21\am\,41.9\as$ & $-03^{\circ}\,21\pr\,47\2pr$ & 1.02   & Optical &  $ 32.2 \pm 2.6$\\
E & WARP J1415.1+3612 & $14\ah\,15\am\,11.1\as$ & $+36^{\circ}\,12\pr\,03\2pr$ & 1.03   & X-ray &  $ 17.2 \pm 2.6$\\
F & ISCS J1432.4+3332 & $14\ah\,32\am\,29.1\as$ & $+33^{\circ}\,32\pr\,48\2pr$ & 1.11   & IR-Spitzer &  $ 20.7 \pm 2.3$\\
G & ISCS J1429.3+3437 & $14\ah\,29\am\,18.5\as$ & $+34^{\circ}\,37\pr\,25\2pr$ & 1.26   & IR-Spitzer &  $ 12.1 \pm 3.1$\\
H & ISCS J1434.5+3427 & $14\ah\,34\am\,28.5\as$ & $+34^{\circ}\,26\pr\,22\2pr$ & 1.24   & IR-Spitzer &  $ 17.8 \pm 2.8$\\
I & ISCS J1432.6+3436 & $14\ah\,32\am\,38.3\as$ & $+34^{\circ}\,36\pr\,49\2pr$ & 1.34   & IR-Spitzer &  $  2.5 \pm 3.2$\\
J & ISCS J1434.7+3519 & $14\ah\,34\am\,46.3\as$ & $+35^{\circ}\,19\pr\,45\2pr$ & 1.37   & IR-Spitzer &  $  2.7 \pm 3.2$\\
K & ISCS J1438.1+3414 & $14\ah\,38\am\,09.5\as$ & $+34^{\circ}\,14\pr\,19\2pr$ & 1.41   & IR-Spitzer &  $ 30.0 \pm 3.1$\\
L & ISCS J1433.8+3325 & $14\ah\,33\am\,51.1\as$ & $+33^{\circ}\,25\pr\,51\2pr$ & 1.5\tablenotemark{a} & IR-Spitzer & $ -2.3 \pm 3.2$\\
M & Cl J1604+4304      & $16\ah\,04\am\,22.6\as$ & $+43^{\circ}\,04\pr\,39\2pr$ & 0.90  & Optical &  $ 10.8 \pm 2.6$\\
N & RCS022056-0333.4  & $02\ah\,20\am\,55.7\as$ & $-03^{\circ}\,33\pr\,10\2pr$ & 1.03   & Optical &  $ 24.4 \pm 2.5$\\
P & RCS033750-2844.8  & $03\ah\,37\am\,50.4\as$ & $-28^{\circ}\,44\pr\,28\2pr$ & 1.1\tablenotemark{a} & Optical & $ 10.9 \pm 2.2$\\
Q & RCS043934-2904.7  & $04\ah\,39\am\,38.0\as$ & $-29^{\circ}\,04\pr\,55\2pr$ & 0.95   & Optical &  $ 17.4 \pm 2.7$\\
R & XLSS J0223.0-0436  & $02\ah\,23\am\,03.7\as$ & $-04^{\circ}\,36\pr\,18\2pr$ & 1.22  & X-ray &  $ 20.1 \pm 2.8$\\
S & RCS215641-0448.1  & $21\ah\,56\am\,42.1\as$ & $-04^{\circ}\,48\pr\,04\2pr$ & 1.07   & Optical &  $ 10.2 \pm 2.1$\\
T & RCS2-151104+0903.3  & $15\ah\,11\am\,03.8\as$ & $+09^{\circ}\,03\pr\,15\2pr$ & 0.97   & Optical &  $  6.8 \pm 2.4$  \\
U & RCS234526-3632.6  & $23\ah\,45\am\,27.3\as$ & $-36^{\circ}\,32\pr\,50\2pr$ & 1.04   & Optical &  $ 17.3 \pm 2.3$\\
V & RCS231953+0038.0  & $23\ah\,19\am\,53.3\as$ & $+00^{\circ}\,38\pr\,13\2pr$ & 0.91   & Optical &  $ 22.6 \pm 2.5$\\
W & RX J0848.9+4452   & $08\ah\,48\am\,56.2\as$ & $+44^{\circ}\,52\pr\,00\2pr$ & 1.26   & X-ray &  $ 11.7 \pm 3.1$\\
X & RDCS J0910+5422   & $09\ah\,10\am\,44.9\as$ & $+54^{\circ}\,22\pr\,08\2pr$ & 1.11   & X-ray &  $ 12.8 \pm 2.0$\\
Y & RDCS J1252.9-2927 & $12\ah\,52\am\,54.4\as$ & $-29^{\circ}\,27\pr\,17\2pr$ & 1.23   & X-ray &  $ 26.3 \pm 2.9$\\
Z & XMMU J2235.3-2557 & $22\ah\,35\am\,20.6\as$ & $-25^{\circ}\,57\pr\,42\2pr$ & 1.39   & X-ray &  $ 14.9 \pm 3.0$
\enddata
\tablenotetext{a}{Photometric redshift}
\tablenotetext{b}{$N_{red}$ is described in Appendix \ref{sec:appendix}}
\tablerefs{A \citep{stanford06a,hilton07a}; B,C \citep{bohringer05a};
D, N, U (Gilbank et al. in prep); E \citep{perlman02a}; F \citep{elston06a}; G, I, J, L \citep{eisenhardt08a};
H \citep{brodwin06a}; K \citep{stanford05a}; M \citep{postman01a}; Q \citep{cain08a};
R \citep{andreon05a,bremer06a}; S \citep{gilbank08a}; V \citep{hicks08a};
W \citep{rosati99a}; X \citep{stanford02a}; Y \citep{rosati04a}; Z \citep{mullis05a}.}
\end{deluxetable*}

%% file: table2.tex
\begin{deluxetable}{lccccc}
\tablewidth{0pt}
%\tabletypesize{\small}
\tabletypesize{\footnotesize}
%\tabletypesize{\scriptsize}
\centering
\tablecaption{\label{table:obstimes} Number of visits and total integration times}
\tablehead{\colhead{ID} & \colhead{$i_{775}$ Visits} & \colhead{$i_{775}$ Int. (s)} & \colhead{$z_{850}$ Visits} & \colhead{$z_{850}$ Int. (s)} & }
\startdata
A & 4 & 3320 & 8   & 16935  \\
B & 8 & 4535 & 9   & 11380  \\
C & 8 & 4110 & 8   & 10940  \\
D & 5 & 2015 & 8   & 13360  \\
E & 7 & 2425 & 7   & 9920  \\
F & 9 & 4005 & 9   & 12440  \\
G & 5 & 2670 & 9   & 15600  \\
H & 5 & 2685 & 9   & 13320  \\
I & 5 & 2235 & 6   & 8940  \\
J & 4 & 1920 & 7   & 11260  \\
K & 5 & 2155 & 7   & 10620  \\
L & 4 & 1845 & 8   & 12280  \\
M & 7 & 2445 & 7   & 10160  \\
N & 6 & 2955 & 9   & 14420  \\
P & 4 & 1560 & 8   & 12885  \\
Q & 4 & 2075 & 9   & 15530  \\
R & 6 & 3380 & 9   & 14020  \\
S & 4 & 2060 & 4   & 5440  \\
T & 5 & 2075 & 5   & 7120  \\
U & 8 & 4450 & 8   & 9680  \\
V & 5 & 2400 & 5   & 6800  \\
W & 3 & 1060 & 6   & 10400  \\
X & 6 & 2250 & 7   & 11000  \\
Y & 3 & 1065 & 6   & 9070  \\
Z & 7 & 8150 & 9   & 14400 
\enddata

\tablecomments{ Observations were modified two months into the survey
to include an $i_{775}$ exposure in each orbit.  Before this time,
targets were observed in only the $z_{850}$ filter.}

\end{deluxetable}

%% file: section4.tex
\subsection{ACS SN Discoveries}

In total, 39 SN candidates were discovered in the ACS images.
A legitimate SN candidate was required to have consistent signal
in each of the four individual exposures in the $z_{850}$ filter
regardless of the reference image that was used.
Several false detections were identified due to bad pixels in reference images.
In addition, several transient objects were determined to be active galactic
nuclei (AGN) due to obvious spectroscopic features.
The 39 SN candidates occurred away from the core of the host
galaxy or were hosted by a galaxy with no spectroscopic evidence of AGN activity.
All candidates that satisfy these criteria
are identified as SNe throughout the remainder of the text.

\begin{figure}[h]
\begin{center}
\includegraphics[scale=0.95]{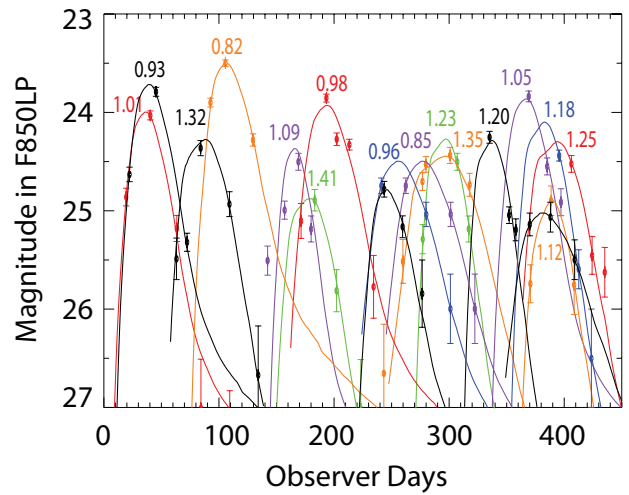}
\end{center}
\caption[SN lightcurves]{\label{fig:lightcurves} Lightcurves and
$z_{850}$ AB magnitudes of the high redshift SNe discovered in the
survey.  Each lightcurve is labeled with the SN redshift in the same
color.  NICMOS and $i_{775}$ observations are not shown, nor are
additional SNe at $z<0.8$.}
\end{figure}

The SNe in the targeted clusters are reported in Table~\ref{table:cluster SNe}
and the SNe discovered in the field are reported in
Table~\ref{table:Field SNe}.  The constant pre-scheduled cadence of
observations provided full lightcurves of most high redshift SNe as
shown in Figure~\ref{fig:lightcurves}.
SN hosts with the properties of red-sequence (or early-type) galaxies
were identified in a preliminary analysis
as described in Appendix~\ref{sec:appendix}.
We obtained at least three
observations with S/N~$>5$ for eight of the nine SNe hosted by
galaxies with color and magnitude near the red-sequence at their
redshift (one other SN was discovered near the edge of the orbital
visibility window).

\input{table3}

\input{table4}

\subsection{Spectroscopic Follow-up}\label{subsec:spec follow}

In this paper, we describe the observations that were taken with FORS1, FORS2 and
DEIMOS. The FOCAS observations are described in Morokuma et al., (2009,
in preparation) and the LRIS observation of the unusual transient 
SN~SCP06F6 is described in \citet{barbary08a}.
				   
\subsubsection{Observations with FORS1 and FORS2}

With the exception of the unusual transient object SN~SCP06F6 \citep{barbary08a},
observations with FORS1 and FORS2 were performed with the 300I grism and the
OG590 order sorting filter. This configuration provides a wavelength range starting at 5900
\AA\ and extending to approximately 10000 \AA\ with a dispersion of 2.3
\AA\ per pixel.
SN~SCP06F6 was observed with the 300V grism covering a wavelength range of
3900 to 8300 \AA\ without an order sorting filter.

Since the observations had to be carried out on short notice (we wanted
the SN to be observed while it was near maximum light), most multislit
observations were done with the multi-object spectroscopic (MOS) mode
of FORS2 rather than with pre-cut masks.
The MOS mode consists of 19 movable slits (with lengths
that vary between 20" and 22") that can be inserted into the focal
plane. The slit width was set to 1".  On one occasion, we used the
long slit mode on FORS1 (cluster D) and on one other occasion, when
the MOS mode was unavailable on FORS2, we used a pre-cut mask (cluster
C).

In observations when the SN was near maximum light, one movable slit was placed on the supernova. The
other movable slits were placed on candidate cluster galaxies or field
galaxies. Later, once the supernova had faded from view, a new mask
was designed. Again, one movable slit was placed on the host without SN
light and other movable slits were placed on candidate cluster
galaxies or field galaxies. Apart from the host of the SN, we did not
re-observe galaxies for which redshifts could be derived from the
spectra that were obtained in the first mask.

Most clusters were observed at least twice and thus have
extensive spectroscopic coverage.
For each setup, between 2 and 9 exposures with exposure times varying
between 600 to 900 seconds were taken. Between each exposure, the
telescope was moved a few arcseconds along the slit direction. These
offsets, which shift the spectra along detector columns, allow one to
remove detector fringes.  The process is described in \citet{hilton07a} where
additional details of data processing are found.

FORS1 and FORS2 were used to observe 14 candidates in 7 clusters.
A list of the clusters that were targeted with FORS1 and FORS2 is given
in Table~\ref{table:SpecFields}.
The total integration times, date of observations, and other details
of the spectroscopic follow-up of clusters D, F, and R are
listed in Table~\ref{table:SpecDetails}. Details for clusters A, C, U
and Z are presented in \citet{hilton07a}, \citet{santos09a},
Gilbank et al. (2009, in preparation), and Rosati et al. (2009, in
preparation), respectively.

\input{table5}
\input{table6}

\subsubsection{Observations with DEIMOS}

DEIMOS was used to observe the hosts of SN candidates in 6 clusters.  A
list of the clusters that were targeted with DEIMOS are given in
Table~\ref{table:SpecFields}.  All observations were done using the
600ZD grating and the OG550 order sorting filter.  This configuration
has a typical wavelength
coverage of 5000 to 10000 \AA\ and a dispersion of 0.64 \AA\ per pixel.

All DEIMOS observations were done with slitmasks containing between 24
and 90 slits with a width of 1''.  In all masks, slits were placed on SN
candidate host galaxies.  Other slits were inserted on candidate
cluster galaxies and field galaxies as space allowed.
Spectroscopically measured redshifts for clusters F
(ISCS~J1432.4+3332), H (ISCS~J1434.5+3427), and K (ISCS~J1438.1+3414)
are presented in \citet{eisenhardt08a}.  Details for cluster R
(XLLS~J0223+0-0436), L (ISCS~J1433.8+3325) and V (RCS~J2319.8+0038)
are presented in Table~\ref{table:SpecDetails}.

For each DEIMOS slitmask three or four exposures with exposure times
varying from 1200s to 1800s were taken.  We reduced the DEIMOS images
with a slightly modified version of the DEEP2 pipeline developed at UC
Berkeley.\footnote{See http://astro.berkeley.edu/$\sim$cooper/deep/spec2d.}
In total, seven SN candidates or host galaxies were observed in six clusters
using the DEIMOS spectrograph.

\subsubsection{Results}

We obtained redshifts for all cluster SNe and all field SNe that were
hosted by galaxies that were near the red-sequence at the
corresponding redshift.  Every SN was targeted for spectroscopic
follow-up with the exception of SN~SCP06Q31, SN~SCP05D55,
SN~SCP06Z52, and SN~SCP06Z53.
The level of completeness in the spectroscopic follow-up
allows us to rule out obvious AGN based on spectroscopic features.
In total, 26 of the 35 supernovae targeted for spectroscopy have
redshifts.  We also obtained redshifts for over 200 cluster galaxies
in a subset of 16 of the 25 clusters that were used to fill the slitmasks
in parallel observations.

\begin{figure*}[h!]
\begin{center}
\includegraphics[scale=0.5,bb=114 150 497 540]{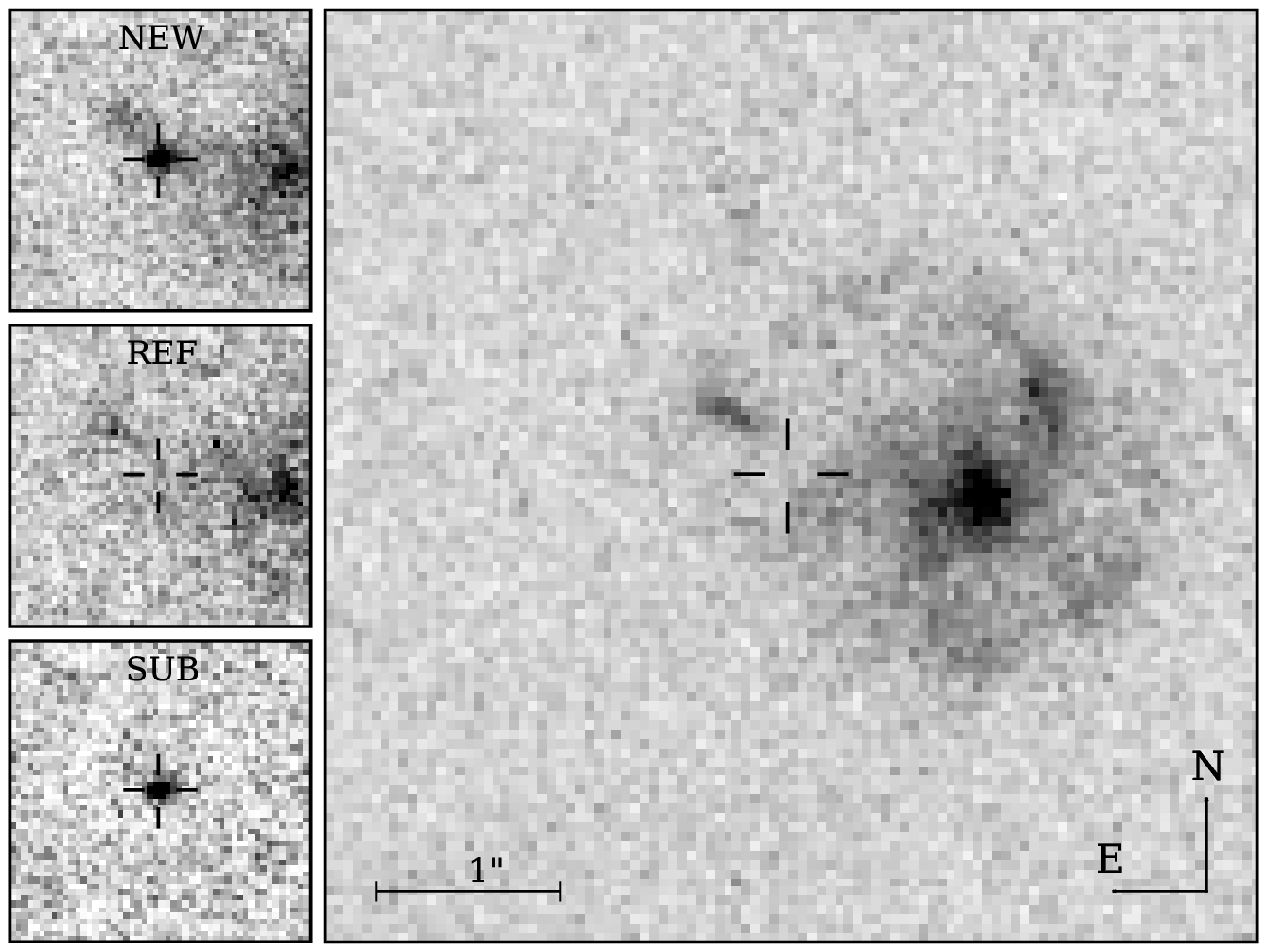}
\includegraphics[scale=0.45,bb=1 160 592 718]{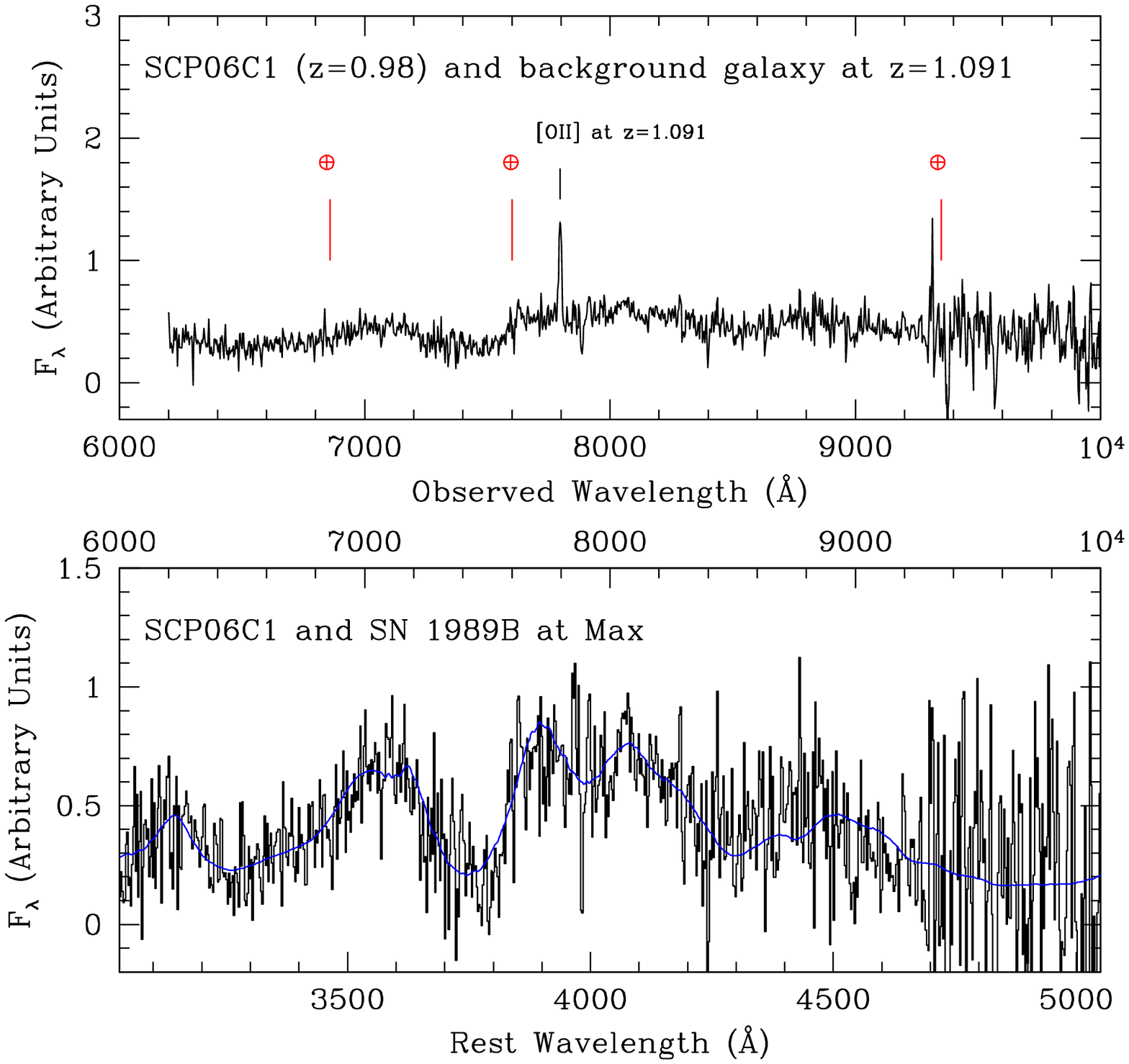}
\end{center}
\caption[SCP06C1]{\label{fig:VLTspectra1}{\bf Left:}
The panels on the left hand side demonstrate the discovery of SN~SCP06C1
from the ACS images
labeled "NEW" (host galaxy with SN light), "REF" (host only), and
"SUB" (residual after subtracting the reference image of the host galaxy).
This SN occurred
along the line-of-sight to a large background galaxy at $z=1.091$.
The main panel shows the much brighter background galaxy
centered west of the SN position and the potential SN host which lies NE of the
SN position but does not have spectroscopic confirmation.
{\bf Right/Upper Panel:} The FORS2 
spectrum of SN~SCP06C1, a secure SN~Ia at $z=0.98$. 
The [OII]\,$\lambda\lambda 3727$ doublet from the background galaxy can be
seen in the spectrum. The spectrum is plotted in the observer frame.
Strong telluric absorption features have been removed.
The residuals are marked with $\oplus$. {\bf
Right/Lower Panel:} After subtracting the spectrum of the background galaxy,
the spectrum is re-binned and plotted in the rest frame. The spectrum
of SN~1989B at maximum light is
the best matching nearby SN Ia.  This SN is plotted at the blue continuous line.}
\end{figure*}

With the exception of SN~SCP06C1, all redshifts were
determined from spectroscopic features of the host galaxy.  
As shown in Figure~\ref{fig:VLTspectra1}, it was
difficult to identify the host galaxy of SN~SCP06C1 in the ACS images
or in the follow-up spectroscopy due to contamination from a much
brighter background galaxy.  The well-measured spectrum of this SN
(see Figure~\ref{fig:VLTspectra1}) clearly identifies this supernova as
a SN~Ia at the redshift of the cluster. 
SN~SCP06C1 is assigned a Confidence Index of 4 for spectroscopic classification.

\begin{figure*}[h!]
\begin{center}
\includegraphics[scale=0.65]{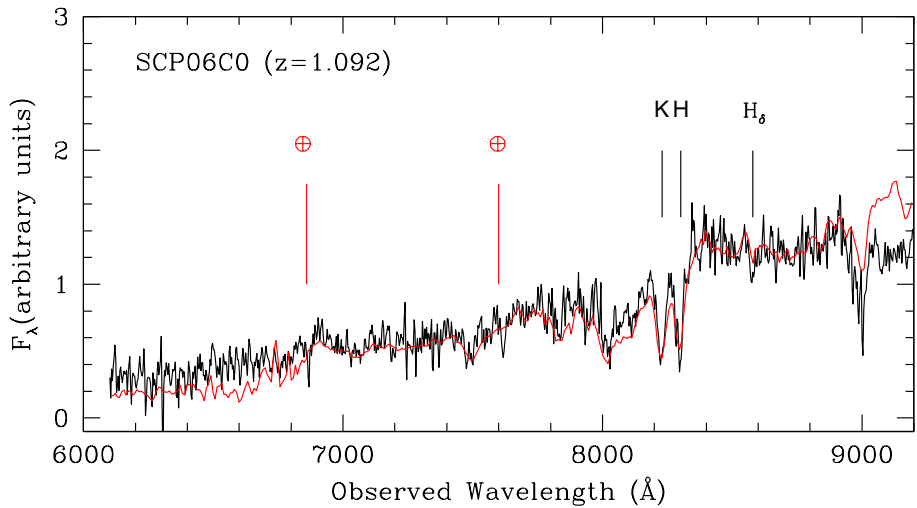}
\end{center}
\vspace{-6cm}
\caption[SCPSpectra]{\label{fig:VLTspectra3} The spectrum of SN~SCP06C0,
an example in which the SN was considerably fainter than the host
galaxy (black line).
The solid red line is the spectral
template of an elliptical galaxy from \citet{kinney96a}. The colors
of the host of SN~SCP06C0 places it within the red sequence of the
cluster (see Appendix A to see how this is defined) and the spectrum is,
qualitatively, a good match to the spectrum of an elliptical galaxy.} 
\end{figure*}

The spectra of the host galaxies were generally clearly indicative of
host environment.  When the photometric and spectroscopic properties
of the host indicated an early-type galaxy (i.e. it landed on the red
sequence and the spectrum lacks obvious nebular emission lines and
stellar Balmer absorption lines), then we could classify the supernova
as a SN  Ia with strong confidence. An example is shown in Figure~\ref{fig:VLTspectra3}.
This typing of SNe Ia in early-type galaxies is well-established at low
redshift \citep{cappellaro99a,hakobyan08a}.
SNe that are hosted by galaxies that appear to be red-sequence galaxies are
identified in a footnote in Table~\ref{table:cluster SNe} and Table~\ref{table:Field SNe}.

\begin{figure*}[h!]
\begin{center}
\includegraphics[scale=0.45]{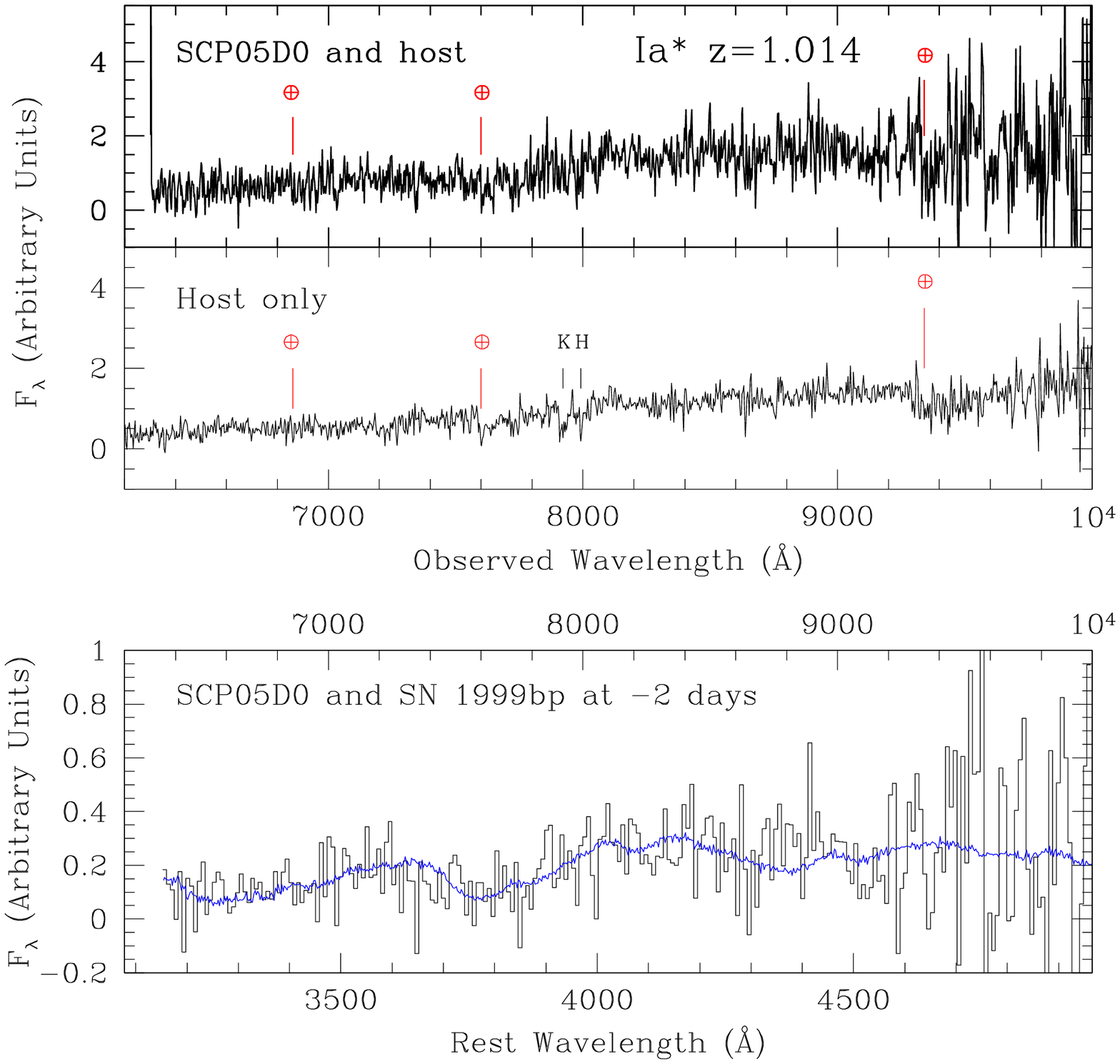}
\includegraphics[scale=0.45]{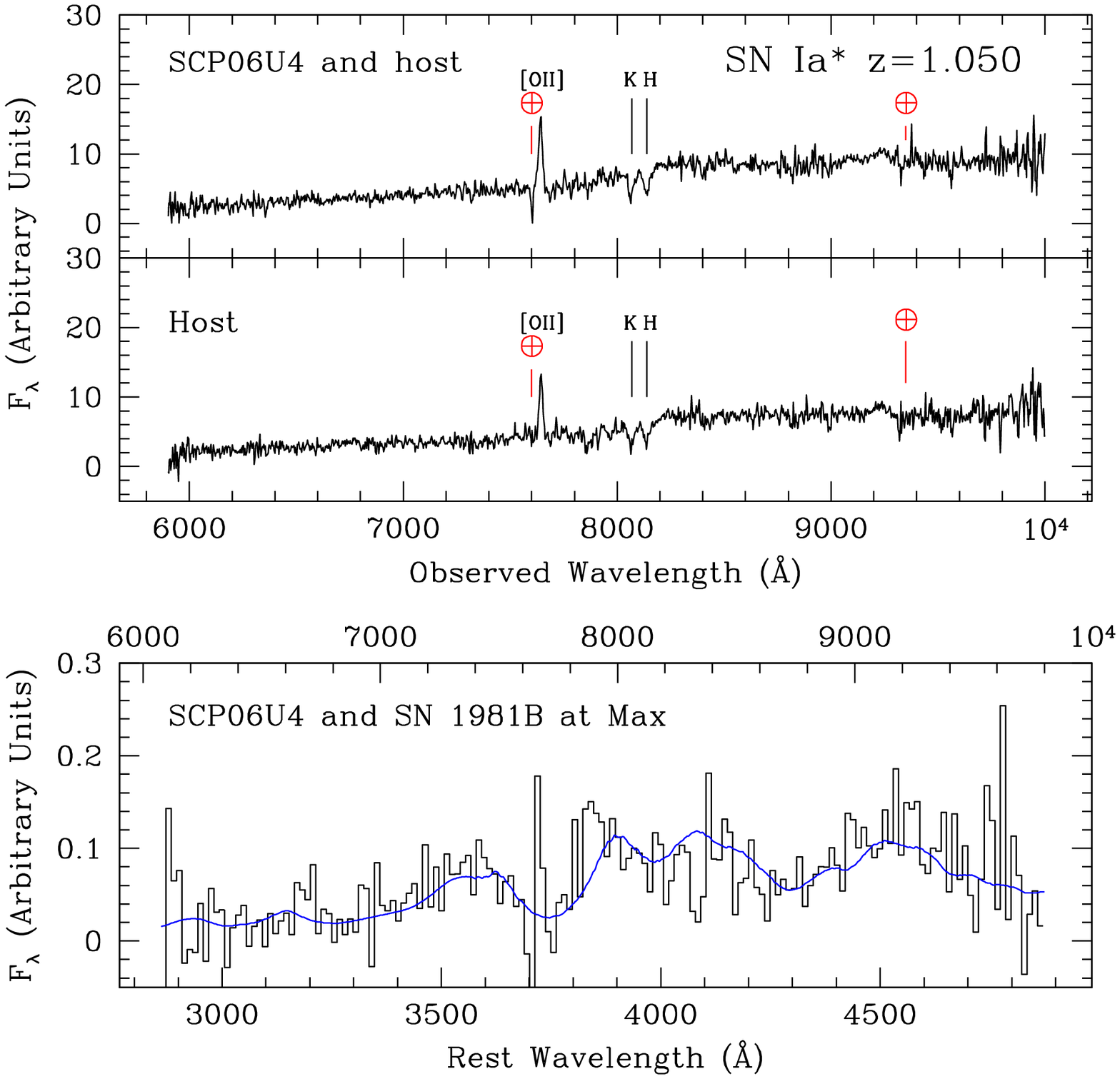}
\end{center}
\caption[SCPSpectra]{\label{fig:VLTspectra2}Spectra of
two SNe that are heavily contaminated by the
light of the host galaxy.
The three panels in each plot show, from top
to bottom, the spectrum of the supernova and the host, the spectrum of
the host without SN light, and the host-subtracted
supernova spectrum in the rest frame.
The best matching nearby SNe~Ia are plotted in the lowest panels in blue.}
\end{figure*}

In several cases, we were able to further confirm SN type at these
high-redshifts by studying the spectrum of the SN itself.
Using the spectrum for supernovae that are hosted by bright early-type
galaxies is a challenging task, as the
hosts can be several magnitudes brighter than the supernovae.  In some
observations, however, we were able to observe
the supernova when it was close to maximum light and when the host was
not too dominant. In these cases
we could identify, positively in some cases and tentatively in others,
the supernova type (see Figure~\ref{fig:VLTspectra1} for an
example). For these SNe we followed the spectroscopic analysis procedure outlined
in \citet{lidman05a}. In quite
a few cases, we re-observed the host when the supernova had faded from
view. By providing a measure
of background contamination at the location of the SN, the second set of
spectroscopic observations thus
facilitates the task of subtracting the host spectrum and identifying the
supernova, as shown for SNe SCP06U4
and SCP05D0 in Figure~\ref{fig:VLTspectra2}.
With a best fit spectroscopic phase of -2 days for SN~SCP05D0 and 0 days for SN~SCP06U4,
the spectral templates are consistent with the best fit epoch of observation
from the SN lightcurves.

We quantify the degree of confidence for classification
of a candidate as a SN Ia using the index described in
\citet{howell05a}.  A Confidence Index ranging from 5 to 0 is
assigned for each SN candidate based on its spectral
features and consistency with the epoch of observation predicted by the best-fit lightcurve.
Candidates with 5, 4, 3, 2, 1, and 0 are {\it certain SN Ia},
{\it highly probable SN Ia}, {\it probable SN Ia}, {\it unknown object},
{\it probably not SN Ia}, and {\it not SN Ia}, respectively.
We report SNe that are typed with a confidence index of 3 or higher
in footnotes to Table~\ref{table:cluster SNe} and Table~\ref{table:Field SNe}.

\clearpage

SN~SCP05D0 and SN~SCP06U4 are both assigned a Confidence Index of 3 for spectroscopic classification,
meaning that both are likely Type Ia and represented in the figure as SN Ia$^*$.
Not shown is the spectrum of SN~SCP06D5 which was observed from VLT
and classified with a Confidence Index of 3.
Similarly classified SNe that were observed with Subaru will be presented
in Morokuma et al. (in preparation).

\subsection{SN Yield}

\begin{figure}[!h]
\begin{center}
\includegraphics[scale=0.6]{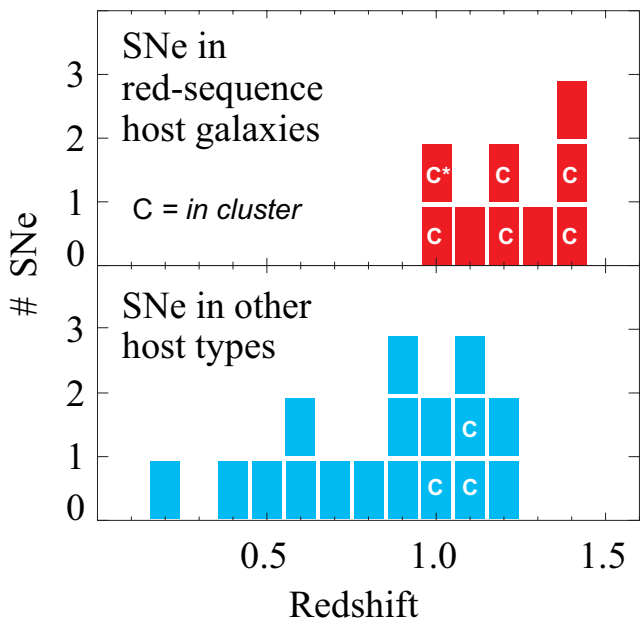}
\end{center}
\caption[SN discoveries]{\label{fig:histogram} The redshift
  distribution of the SNe discovered in our ACS program.  The upper
  panel shows the distribution of red-sequence galaxies that had SNe Ia,
  with the cluster members indicated by ``C''. The asterisk indicates
  [OII] emission in the spectrum of a galaxy with red-sequence color
  and spectral features as explained in the text.  The
  lower panel shows the distribution for the other SN hosts.
  For $z > 0.95$, the number of SNe in cluster galaxies (9)
  is comparable to the number of SNe in field galaxies (7).}
\end{figure}

One of the primary goals of the HST Cluster SN Survey was to increase
the yield of high redshift SNe relative to previous HST surveys.  With
nine of the 16 SNe we identified at $z>0.95$ coming from galaxy clusters, the
program demonstrates that we more than double the number of high
redshift SNe by searching in rich cluster fields as shown in
Figure~\ref{fig:histogram}.  We obtain an even greater enhancement of
the number of SNe in red-sequence host galaxies.  Six of the nine SNe
hosted by red-sequence galaxies came from clusters at $z > 0.95$.
These SNe offer an unbiased measure of
intrinsic color and of SN properties in a specific host environment as
outlined in the survey strategy.

Each SN can be checked for dust by analyzing the host properties
relative to the cluster red-sequence. The host
of SN~SCP06U4 (Figure~\ref{fig:VLTspectra2}) is an interesting case. 
Here we observe
[OII] emission in the spectrum of a galaxy with otherwise early-type
color, morphology, and spectral features.  There is evidence
\citep{yan06a} that those features may be coming from
quiescent red galaxies, with [OII] emission tracing
AGN activity rather than star formation.
After further tests of morphology, color, and spectral properties of
the host of SN~SCP06U4 relative to the
galaxies that comprise the red-sequence of RCS234526-3632.6,
this SN may yet be included in the subsample of dust-free environments.

For comparison, the SN searches in the GOODS fields were not centered on rich clusters
and revealed a much lower yield of SNe hosted by early-type galaxies.
In the GOODS SN program, 360 HST orbits were used in the initial search and more than 390 additional orbits were
used to obtain full lightcurves and spectral confirmation of 23 SNe at $z>1$.
Six SNe with early-type hosts were discovered at $z>0.9$ \citep{riess07a}.
Scaling the yield of SNe from our 219-orbit program, we would expect to find 27
red-sequence hosted SNe with complete lightcurves in the same total number of orbits
as the GOODS program in the same redshift range.
If we only consider the ACS search visits for the two programs (360 vs 188),
we would expect to discover 15 red-sequence hosted SNe with complete lightcurves
in fields with rich clusters as opposed to the six that were discovered in the
GOODS survey.
In the following section we will discuss the boost in SNe from an astrophysical
perspective, here we explain the additional gain due to the efficiency
of the survey strategy.

There are several reasons for the improved efficiency from this new
observation technique.
For one, the density enhancement of galaxies in the cluster fields provides
more potential $z>0.9$ SN hosts than in a blind search.
This increases the percentage of orbits in the pre-scheduled observations
with active SN light - 67 of the 188 ACS orbits
were effectively used to measure SNe lightcurves in clusters compared
to 54 orbits for SNe hosted by $z>0.95$ field galaxies.
The shorter amount of time between observations of a given cluster field also 
increases the likelihood that a SN is discovered early in its lightcurve.
Early discovery enables the scheduling of a single high signal-to-noise NICMOS
observation near maximum light for constraints on color for SNe at $z<1.3$
and a full lightcurve of NICMOS observations for SNe at higher redshift.
Finally, the SNe can be characterized by the evolutionary history
of their red-sequence host galaxies.  SNe Ia can therefore be identified with
similar confidence to high redshift SNe from previous programs without difficult
direct spectroscopic typing.

%% file: table3.tex
\begin{deluxetable*}{lccccc}
\centering
\tablewidth{0pt}
%\tabletypesize{\small}
\tabletypesize{\footnotesize}
%\tabletypesize{\scriptsize}
\tablecaption{\label{table:cluster SNe} SNe hosted by spectroscopically confirmed cluster members}
\tablehead{\colhead{SN Name} & \colhead{SN} & \colhead{Cluster} & \colhead{Galaxy} & \colhead{Distance from} & \colhead{Distance from}
\\
\colhead{} & \colhead{Nickname} & \colhead{ID} & \colhead{Redshift (z)} & \colhead{Center (kpc)} & \colhead{Center ($\2pr$)}
}
\startdata
SN SCP06B4 & Michaela & B  &  1.117 &     $143$  & $17.4$   \\
SN SCP06C1\tablenotemark{a} & Midge &    C  &  0.98 &      $471$  & $58.8$   \\
SN SCP05D0\tablenotemark{a,b} & Frida &    D  &  1.015 & $47$   & $6.1 $   \\
SN SCP06F12 & Caleb &        F  &  1.110 &      $407$  & $49.4$   \\
SN SCP06H5\tablenotemark{b} & Emma &     H  &  1.231 & $111$  & $13.3$    \\
SN SCP06K0\tablenotemark{b} & Tomo &     K  &  1.415 & $20$   & $2.4 $   \\
SN SCP06K18\tablenotemark{b} & Alexander & K & 1.411 & $791$  & $93.1$    \\
SN SCP06R12\tablenotemark{b} & Jennie &  R  &  1.212 & $428$  & $51.3$    \\
SN SCP06U4\tablenotemark{a,b} & Julia &    U  &  1.050 &     $318$  & $39.1$     \\
\enddata
\tablenotetext{a}{Spectroscopy indicates likely SN Ia}
\tablenotetext{b}{Denotes host galaxy with color and magnitude consistent with
the cluster red-sequence}
\end{deluxetable*}

%% file: table4.tex
\begin{deluxetable}{lccc}
\centering
\tablewidth{0pt}
%\tabletypesize{\small}
\tabletypesize{\scriptsize}
\tablecaption{\label{table:Field SNe} SNe hosted by field galaxies and galaxies with unknown redshift}
\tablehead{\colhead{SN} & \colhead{SN} & \colhead{Cluster} & \colhead{Galaxy}\\
\colhead{Name} & \colhead{Nickname} & \colhead{ID} & \colhead{Redshift (z)}}
\startdata
%\cutinhead{$z>0.95$}
\multicolumn{4}{c}{$z>0.95$}\\
\hline
SN SCP06A4\tablenotemark & Aki &    A  &  1.193 \\
SN SCP06C0\tablenotemark{b} & Noa &      C  &  1.092 \\
SN SCP05D6\tablenotemark{b} & Maggie &   D  &  1.315 \\
SN SCP06G3 & Brian &    G  &  0.962  \\
SN SCP06G4\tablenotemark{a,b} & Shaya &    G  &  1.350 \\
SN SCP06N33 & Naima &   N  &  1.188  \\
SN SCP06T1 & Jane &         T  &  $1.112$ \\
%\cutinhead{$z<0.95$ or redshift unknown}
\hline
\multicolumn{4}{c}{$z<0.95$ or redshift unknown}\\
\hline
SN SCP06B3 & Isabella & B  &  0.744  \\
SN SCP06C7 & Onsi &         C  &  0.606  \\
SN SCP05D55 &  &         D  &  $--$  \\
SN SCP06E12 &  &        E  &  $--$  \\
SN SCP06F6\tablenotemark{c} &  &  F  &  $--$ \\
SN SCP06F8 & Ayako &    F  &  $--$ \\
SN SCP06H3\tablenotemark{a} & Elizabeth& H  &  0.851 \\
SN SCP06L21 &  &        L  &  $--$ \\
SN SCP05N10 & Tobias &  N  &  0.203  \\
SN SCP06N32 &  &        N  &  $--$   \\
SN SCP05P1 & Gabe &     P  &  0.926  \\
SN SCP05P9\tablenotemark{a} & Lauren &   P  &  0.821  \\
SN SCP06Q31 &  &        Q  &  $--$  \\
SN SCP06U2 & William &         U  &  0.543 \\
SN SCP06U6 &  &         U  &  $--$ \\
SN SCP06U7 & Ingvar &   U  &  0.892 \\
SN SCP06X18 &  &        X  &  $--$ \\
SN SCP06X26 & Joe &        X  &  $--$ \\
SN SCP06X27 & Olivia &  X  &  0.435 \\
SN SCP06Z5\tablenotemark{a} & Adrian &   Z  &  0.623 \\
SN SCP06Z13 &  &        Z  &  $--$ \\
SN SCP06Z52 &  &        Z  &  $--$ \\
SN SCP06Z53 &  &        Z  &  $--$
\enddata
\tablenotetext{a}{Spectroscopy indicates likely SN Ia}
\tablenotetext{b}{Denotes host galaxy with color and magnitude consistent with
the red-sequence at the galaxy redshift}
\tablenotetext{c}{Unusual SN-like transient described in \citet{barbary08a}}
\end{deluxetable}

%% file: table5.tex
\begin{deluxetable*}{llll}
  \centering
  \tablewidth{0pt}
  \tabletypesize{\scriptsize}
  \tablecaption{\label{table:SpecFields} Spectroscopically targeted SNe and hosts}
  \tablehead{\colhead{ID} & \colhead{Cluster} & \colhead{Instrument} & \colhead{SN Names}}
  \startdata
  A & XMMXCS J2215.9-1738 & FORS2   & SN SCP06A4 \\ 
  C & XMMU J1229.4+0151   & FORS2   & SNe SCP06C0, SCP06C1, SCP06C7 \\ 
  D & RCS J0221.6-0347    & FORS1/2 & SNe SCP05D0, SCP05D6 \\ 
  F & ISCS J1432.4+3332   & FORS2   & SN SCP06F6\tablenotemark{a} \\ 
  R & XLSS J0223.0-0436   & FORS2   & SN SCP06R12 \\ 
  U & RCS J2345.4-3632    & FORS2   & SNe SCP06U2, SCP06U4, SCP06U6\tablenotemark{a}, SCP06U7 \\ 
  Z & XMMU J2235.2-2557   & FORS2   & SNe SCP06Z5, SCP06Z13\tablenotemark{a}  \\ 
  F & ISCS J1432.4+3332   & DEIMOS  & SN SCP06F12           \\
  H & ISCS J1434.4+3426   & DEIMOS  & SN SCP06H5            \\
  K & ISCS J1438.1+3414   & DEIMOS  & SNe SCP06K0, SCP06K18 \\
  L & ISCS J1433.8+3325   & DEIMOS  & SN SCP06L21\tablenotemark{a}      \\
  R & XLSS J0223.0-0436   & DEIMOS  & SN SCP06R12           \\
  V & RCS J2319.8+0038    & DEIMOS  & SN SCP06V6\tablenotemark{b}            \\
  \enddata
\tablenotetext{a}{Redshift undetermined}
\tablenotetext{b}{Spectroscopically confirmed as AGN}
\end{deluxetable*}

%% file: table6.tex
\begin{deluxetable*}{llllllll}
\centering
\tablewidth{0pt}
\tabletypesize{\scriptsize}
\tablecaption{\label{table:SpecDetails} Observing details for spectroscopic follow-up}
\tablehead{\colhead{ID} & \colhead{Instrument} & \colhead{Mode} & \colhead{Slits} & \colhead{Grating or Grism/Filter} & \colhead{Exposure times} & \colhead{Air Mass} & \colhead{Date (UT)}}
\startdata
D   &   FORS1   &  LSS  &     1  &     300I/OG590   &            7x750    &       1.1  &           17/08/2005\\
D   &   FORS2   &  MOS  &     13 &     300I/OG590   &            6x900    &       1.1  &           19/12/2005\\
R   &   FORS2   &  MOS  &     12 &     300I/OG590   &            4x700    &       1.1  &           31/08/2006\\
R   &	FORS2   &  MOS  &     12 &     300I/OG590   &            6x900    &       1.1  &           30/09/2006\\
F   &   FORS2   &  MOS  &     19 &     300V         &            2x600    &       2.1  &           20/05/2006\\
R   &   DEIMOS  &  MOS  &     45 &     600ZD/OG550  &           3x1200    &       1.1  &           28/08/2006\\
R   &   DEIMOS  &  MOS  &     42 &     600ZD/OG550  &           4x1800    &       1.3  &           28/08/2006\\
R   &   DEIMOS  &  MOS  &     36 &     600ZD/OG550  &           4x1800    &       1.1  &           13/10/2007\\
V   &   DEIMOS  &  MOS  &     24 &     600ZD/OG550  &           4x1800    &       1.1  &           28/08/2006\\
\enddata
\end{deluxetable*}

%% file: section5.tex
The results obtained with this SN survey show qualitative and quantitative
improvements in the SN yield per orbit over a SN survey targeting random fields.
It is also important
to determine if they are consistent with the
improvement expected from the overdensity of red-sequence galaxies in
the targeted clusters.  Here we compare the distribution of cluster
SNe and the increased discovery rate to the distribution and
overdensity of red-sequence cluster members\footnote{It should be
noted that the analysis here is a specific measure of the richness of
red-sequence galaxies to illustrate the benefits of the SN survey
strategy and does not attempt to quantify the additional blue-galaxy
cluster population.}.  

\begin{figure*}[t!]
\begin{center}
\centerline{
  \includegraphics[scale=0.35]{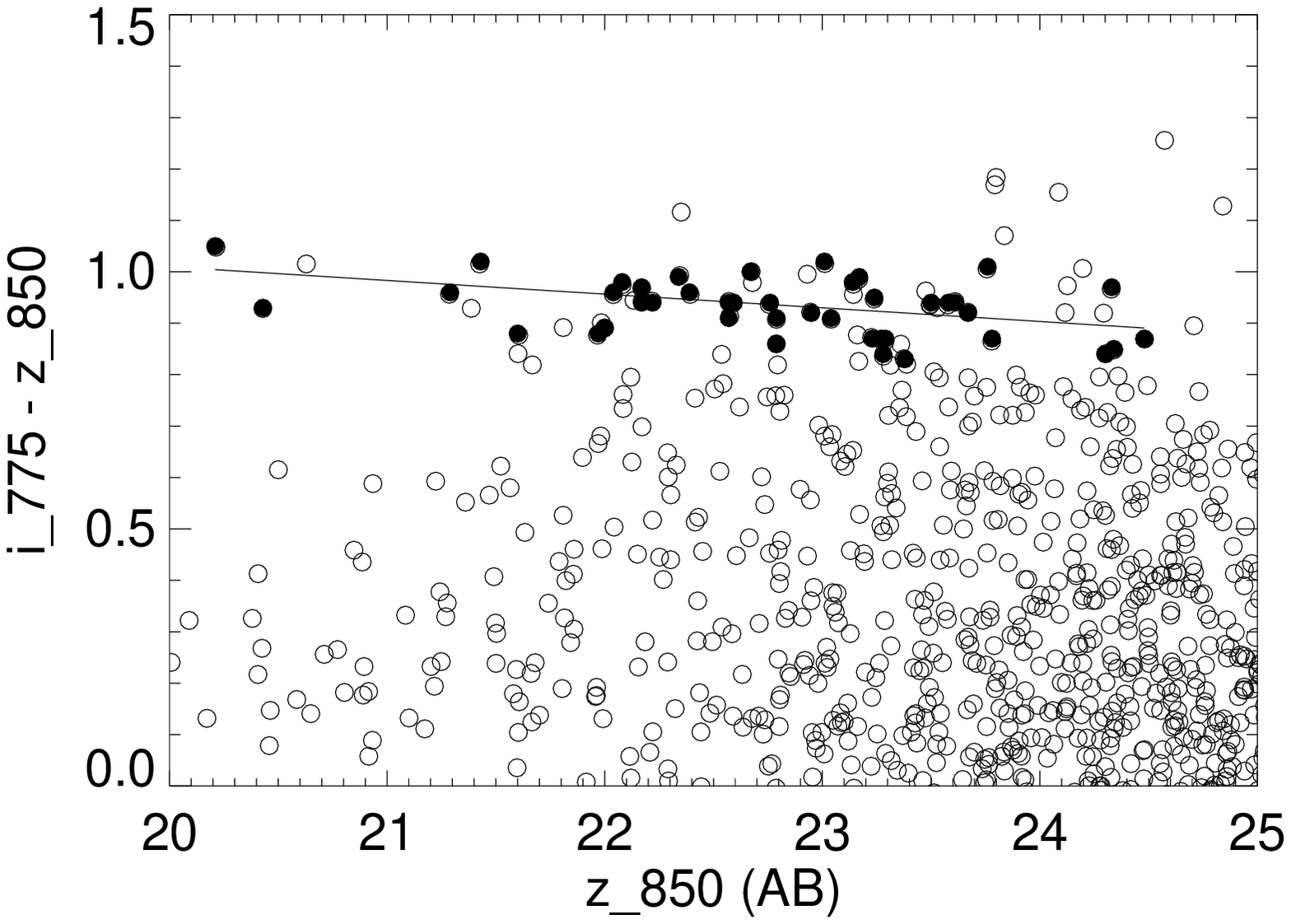}
  \includegraphics[scale=0.35]{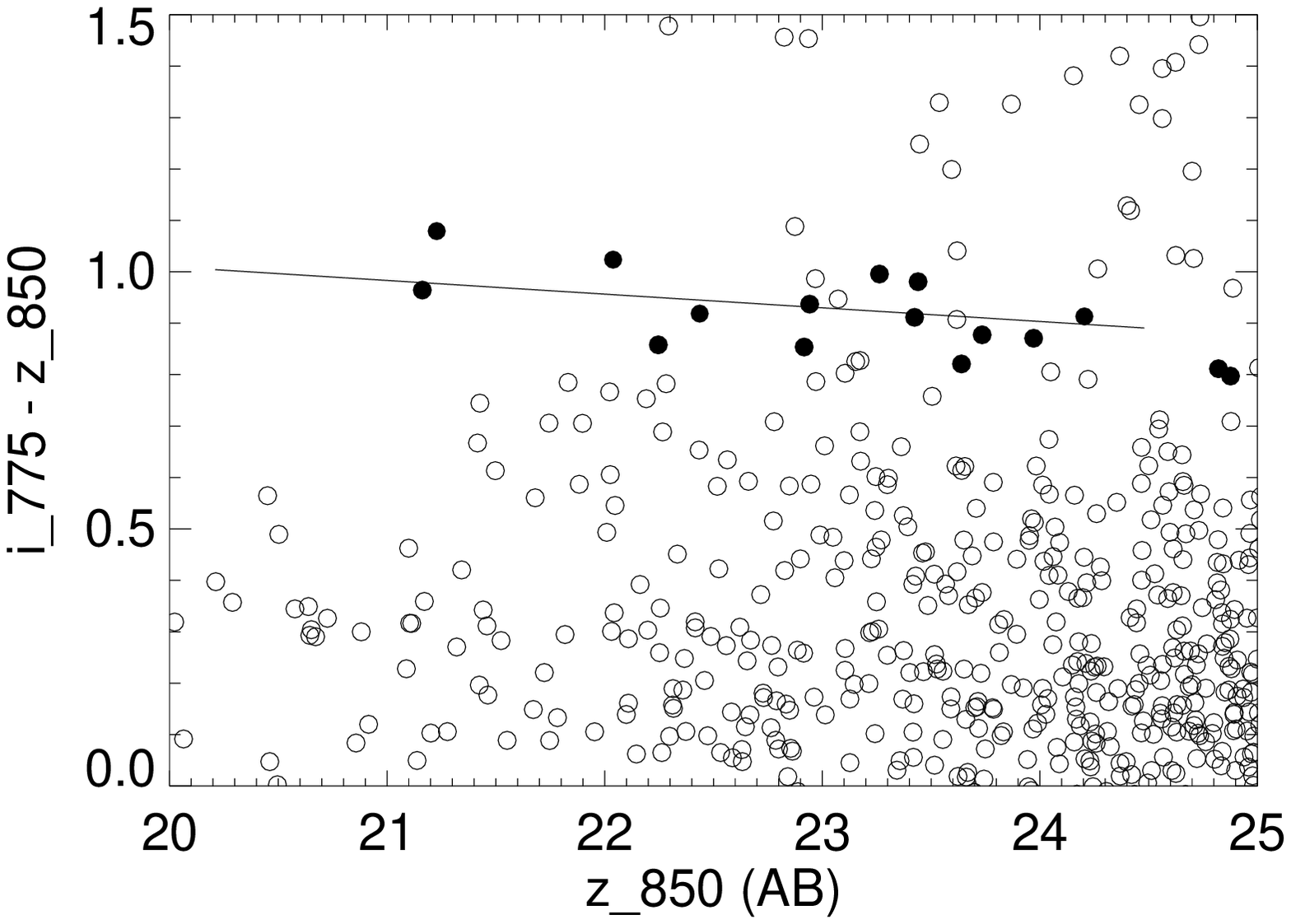}
  \includegraphics[scale=0.35]{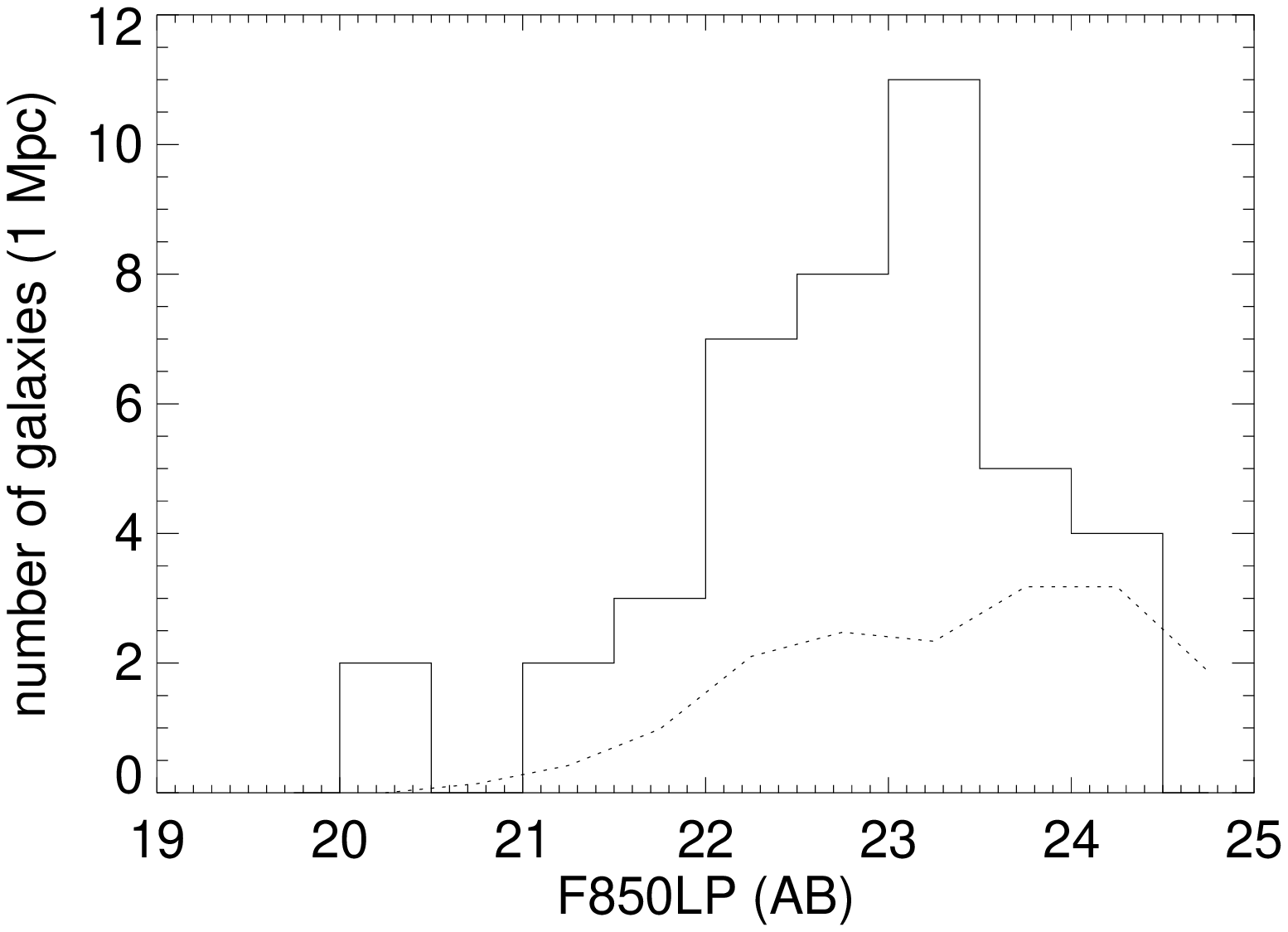}
}
\end{center}
\caption[Color Magnitude Diagram]{\label{fig:cmd}
Red-sequence of WARP J1415.1+3612 compared to field galaxies from the
GOODS data.  {\bf (a) Left:} The color magnitude diagram for galaxies
in the image of WARP J1415.1+3612.  {\bf (b) Center:} The color
magnitude-diagram for galaxies in a random GOODS field of the same
area.
In both color magnitude diagrams, the solid circles represent galaxies
with a color and morphology expected for an early type cluster member,
the open circles represent all galaxies in the image, and the solid line
illustrates the modeled red-sequence at $z=1.03$.  Some objects represented by open
circles had colors and magnitudes near the red-sequence but did not have
morphologies consistent with an elliptical galaxy.
Other objects had a morphology consistent an elliptical galaxy but did not
lie near the model red-sequence.
{\bf (c) Right:}
Histogram of the red-sequence members and interlopers.  The solid line
represents galaxies within 0.1 mag of the red-sequence in WARP
J1415.1+3612 and the dashed line represents an estimate of interlopers
derived from the GOODS fields. }
\end{figure*}

We compute a measure of the red-sequence richness, $N_{red}$,
following the approach outlined in Appendix~\ref{sec:appendix}.
The parameter $N_{red}$ serves as a proxy for the total stellar mass
in cluster-elliptical galaxies, and thus provides the relative density comparisons.
In another analysis (Barbary et al., in prep) we address the more 
detailed questions required to compare SN production at high
and low redshifts, and place constraints on the evolution
of delay times for SNe Ia in early-type cluster galaxies.
This further analysis will include the precise determination of total 
cluster luminosity, SN detection
efficiency, and the window of time probed during the survey.

The red-sequence richnesses and $68\%$ uncertainties are found in
Table~\ref{table:clusters}.  As an example of the method, we show the
results for the $z=1.03$ cluster WARP J1415.1+3612 (cluster ``E'' in
Table~\ref{table:clusters}) in Figure~\ref{fig:cmd}.
The red-sequence galaxies appear as a clear
overdensity for this cluster relative to the HST GOODS
\citep{giavalisco04a} fields.

The mean red-sequence richness of this sample of clusters computed
within a 1 Mpc radius field of view is $N_{red}=15.9$.
This average red-sequence richness is a factor of 1.2 higher than the number of field galaxies that
would be identified as $0.9 < z < 1.5$ red-sequence galaxies in an arbitrary ACS observation.
We therefore expect roughly 1.2 SNe hosted by cluster red-sequence galaxies for every
SN hosted by a random $z>0.9$ red-sequence galaxy,
or 2.2 times as many red-sequence hosted SNe
in this HST program compared to an HST survey of random fields.

The $N_{red}$ parameter serves as a proxy for cluster richness.
For comparison, we computed the total luminosity
of the cluster red-sequence galaxies relative to the field galaxies
using MAG\_AUTO \citep{bertin96a} as a first order estimate of total light.
The overdensity in cluster luminosity relative to the field was found to be
more significant than the overdensity represented by the $N_{red}$ parameter,
indicating that the $N_{red}$ parameter underestimates the cluster luminosity.
Considering that the average cluster red-sequence galaxy in this sample is brighter
than the average interloper, and given the fluctuations expected from Poisson statistics,
the ratio of six cluster red-sequence SNe to three field red-sequence
SNe discovered in the program may be consistent with the increase that would be expected.
It is also possible that the increase is due to a true enchancement of the SN Ia
rate in cluster red-sequence galaxies relative to field red-sequence
galaxies, as recently found at low redshift by \citet{mannucci08a}. 
A more detailed determination of SN yield with cluster luminosity will be
reported in the aforementioned analysis of SN rates in galaxy clusters.

\begin{figure}[!h]
\begin{center}
\includegraphics[scale=0.5]{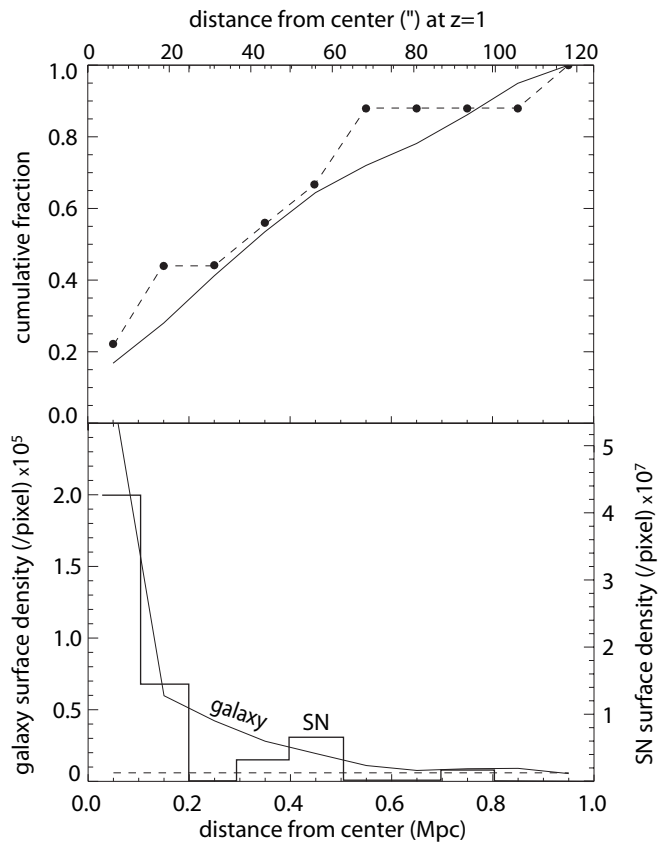}
\end{center}
\caption[Composite radial profile]{\label{fig:radprofile}
{\bf (a) Upper Panel:}
Cumulative distribution of red-sequence galaxies in all 25 clusters with
21 $< z_{850} <$ 24 selected (solid line) and SNe (dashed
line and solid circles).
{\bf (b) Lower Panel:}
The solid line represents the surface density
as a function of distance from the cluster center of galaxies with 21
$< z_{850} <$ 24 selected to lie on the red-sequence for the complete
sample of 25 clusters in this program.  The dashed line represents the
surface density of galaxies satisfying the same criteria in the GOODS
data.  The histogram represents the surface density of cluster SNe
from this program.  Because some of the cluster volume may lie outside the field,
surface densities are computed from the number of pixels
in the stacked ACS images within a distance $r$ from the cluster center.
}
\end{figure}

The radial distribution of SNe discovered in clusters is shown in
Table~\ref{table:cluster SNe}.  Although only $27.7\%$ of the ACS
field of view lies within $1\pr$ of the cluster center, eight of the
nine cluster hosted SNe were found inside this radius.  The high
density of SN events near the cluster core traces the
distribution of red-sequence galaxies as shown in
Figure~\ref{fig:radprofile}.

Applying a Kolmogorov-Smirnov (K-S) statistic analysis to the cumulative distribution yields a maximum
separation $D=0.251$.  The probability of obtaining this separation or larger in
a sample of nine SNe is approximately $55\%$, demonstrating consistency
between the two distributions.  Examination of the increase in yield
and distribution of SNe therefore leads us to conclude that the
strategy of targeting galaxy clusters improved the survey efficiency
with HST as expected.

\begin{figure}[!h]
\begin{center}
\includegraphics[scale=0.42]{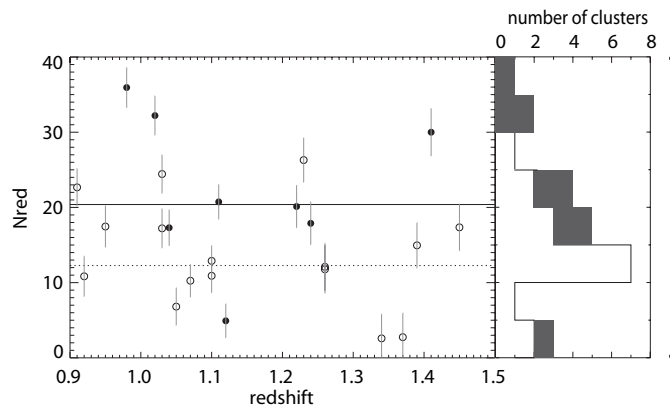}
\end{center} 
\caption[SN vs richness]{\label{fig:snvsrich}
Red-sequence richness as a function of redshift.  Clusters that hosted
supernovae are indicated with filled symbols. Clusters that did not
host supernova are indicated with open symbols. The solid and dotted
lines mark the respective median richnesses of the two samples. The
histogram represents the distribution of richnesses for clusters that
did not host a SN (open) and for clusters that did (filled).}
\end{figure}

A further dramatic result is shown in Figure~\ref{fig:snvsrich} -
the {\it{richer}} clusters were primarily responsible for the
increased productivity.  The median red-sequence richness of clusters
producing a SN was $N_{red} = 20.4$ while the median for clusters not
producing a SN was $N_{red} = 12.1$.

\clearpage

%% file: section6.tex
The strategy of targeting high redshift galaxy clusters results in a
significant boost in the yield of high redshift SNe.  The increased
yield from cluster SNe is consistent with expectations from the
overdensity of red-sequence galaxies.  Nine of the 16 SNe at $z>0.95$
were discovered in galaxy clusters and six of the nine SNe hosted by
red-sequence galaxies were discovered in the cluster environment.  By
doubling the number of total SNe and tripling the number of SNe hosted
by red-sequence galaxies, the new technique of targeting massive
clusters with HST has met its first measurable goal.

The cosmological analysis will be presented in a separate paper with
SN lightcurves and a new high redshift Hubble diagram.
The new sample will be large enough to be used together with
a similar sample of low and moderate redshift SNe to
begin to study the implications of host galaxy environment with
respect to SN properties and derived cosmology.  In addition, the SNe in this
survey will provide improved constraints on SN production in high
redshift clusters compared to previous studies \citep{gal-yam02a,sharon07a,mannucci08a,graham08a}.
Finally, these deep ACS images also enable key studies of the evolution of
massive galaxies and clusters, the measurement of cluster masses via
weak lensing, and an entire program of cluster studies.  Several
related parallel analyses have already been submitted or published
\citep{melbourne07a,hilton07a,eisenhardt08a,barbary08a,hilton09a,santos09a}
using data from this program.

The results of this program are very promising for SN surveys using
the next generation of HST instruments like the IR channel of WFC3.
The WFC3 IR field of view is $60\%$ smaller than ACS but still would
have captured eight of the nine of the cluster SNe discovered in the ACS
program.  A SN survey targeting the same sample of galaxy clusters
with WFC3 would thus be expected to produce nearly the same
yield of cluster hosted SNe.  On the other hand, a conventional SN
survey that does not target clusters would suffer a drop in yield
proportional to the loss of survey area.  More massive clusters at
redshifts out to nearly $z=1.4$ have been discovered in recent surveys that were not
available when the HST Cluster SN Survey took place
\citep[e.g.][]{eisenhardt08a,wilson09a,muzzin09a}.  Since almost all
cluster SNe were found in the richest of the clusters targeted with
ACS, a future survey targeting these newly discovered rich clusters
would lead to an even further increase in the rate of SNe per orbit.
A WFC3 survey would also provide better sensitivity in the NIR,
stretching SN studies to higher redshifts with even better
signal-to-noise than was achieved with ACS.  Such a program would
yield a SN in every cluster, bringing the technique to its logical
conclusion as every HST observation would be used to constrain the SN
lightcurve in a pre-scheduled search with no additional follow-up
imaging.

We thank Julien Guy, Pierre Astier, and Josh Frieman for stimulating
discussion of SN color and host type from their experience in the
analyses of the large SNLS and SDSS-II SN data sets.  
We thank the Aspen Center for Physics for its summer 2007 hospitality during the
preparation of this paper.
Financial support for this work was provided by NASA through program GO-10496
from the Space Telescope Science Institute, which is operated by AURA,
Inc., under NASA contract NAS 5-26555.  This work was also supported
in part by the Director, Office of Science, Office of High Energy and
Nuclear Physics, of the U.S. Department of Energy under Contract
No. AC02-05CH11231, as well as a JSPS core-to-core program
``International Research Network for Dark Energy" and by JSPS research
grant 20040003.  
Support for MB was provided by the W. M. Keck Foundation.
The work of SAS was performed under the auspices of the
U.S. Department of Energy by Lawrence Livermore National Laboratory in
part under Contract W-7405-Eng-48 and in part under Contract
DE-AC52-07NA27344.
The work of PE, JR, and DS was carried out at the Jet Propulsion
Laboratory, California Institute of Technology, under a contract with NASA.

Subaru observations were collected at Subaru Telescope, which is
operated by the National Astronomical Observatory of Japan.  Some of
the data presented herein were obtained at the W.M. Keck Observatory,
which is operated as a scientific partnership among the California
Institute of Technology, the University of California, and the
National Aeronautics and Space Administration.  The authors wish to
recognize and acknowledge the very significant cultural role and
reverence that the summit of Mauna Kea has always had within the
indigenous Hawaiian community.  We are most fortunate to have the
opportunity to conduct observations from this mountain.  Some of
spectroscopic data presented in this paper were taken at the Cerro
Paranal Observatory as part of ESO programs 171.A-0486, 276.A-5034,
077.A-0110 and 078.A-0060.

%% file: appendix.tex
To compute the red-sequence richness of clusters, we used
spectroscopically confirmed early-type galaxies from the observations
discussed briefly in \S\ref{subsec:spec follow} to model the
red-sequence.  We identified the subsample of early-type galaxies with
no significant [OII] emission and a redshift determined from the Ca II
H \& K absorption lines.  In total, 64 early-type galaxies from these
observations were used to develop an empirical model for the
red-sequence for the sample of 25 clusters at these high redshifts.

We used only the photometry from the deep $i_{775}$ and $z_{850}$ ACS
images to generate a model for the red-sequence.  The photometry was
performed using the SExtractor \citep{bertin96a} analysis package.
The analysis of 19 early-type galaxies from RDCS
J1252.9-2927 \citep{demarco07a} and 14 from XMMU J1229.4+0151 produced
an observed RMS scatter from the empirically derived red-sequence of
$0.043$ mag and $0.048$ magnitudes respectively.  The photometry here
was developed to classify potential red-sequence SN hosts and was not optimized to
measure the intrinsic dispersion, resulting in a slightly larger RMS
scatter in RDCS J1252.9-2927 than reported in \citet{blakeslee03a}.

Assuming a 2.5 Gyr stellar population, simple stellar populations of
varying metallicity were used to generate synthetic elliptical galaxy
templates \citep{bruzual03a} for the K-corrections.  For each galaxy,
we chose a template with metallicity that produced the observed
$i_{775}$-$z_{850}$ color and scaled that template to match the
observed $z_{850}$ magnitude.  These spectral templates were
redshifted and passed through the ACS response curves \citep{mack07a}
to generate synthetic $i_{775}$ and $z_{850}$ photometry for each
cluster.  The synthetic $i_{775}$ and $z_{850}$ photometry was used to
populate a color magnitude diagram for each cluster after correcting
for luminosity distance.

Photometric catalogs were generated with the same SExtractor
parameters applied to all objects in the cluster fields that were used
to obtain the photometry of the 64 galaxies that modeled the
red-sequence.  Objects with stellar or disk-like morphologies
identified by CLASS\_STAR $> \, 0.7$, ELLIPTICITY $> \, 0.5$, or
FWHM\_IMAGE $< \, 3.5$ pixels were excluded.  For each ACS field, the
remaining galaxies in the catalog were compared to the model
red-sequence at the redshift of the target cluster.  All galaxies with
a $i_{775}$-$z_{850}$ color within 0.1 mag of the model red sequence
were identified as likely elliptical cluster members.  This criteria
was also used to identify SN hosts as ``red-sequence'' galaxies.  The
red-sequence at the host redshift was used as the template for cluster
and non-cluster SNe.

Many galaxy clusters host a bright cD galaxy that is often used to define the cluster
center. However, several clusters in this sample have two bright central galaxies or no clear
cD galaxy at all. We therefore introduce a simple algorithm to define the center of a cluster.
First, we computed a rough centroid determined by the average position of all galaxies
with color and luminosity consistent with the model red-sequence. Next, we selected the
five brightest galaxies within 0.3 - 0.6 Mpc of the rough center, where the exact aperture
depended on the number of red-sequence galaxies and compactness of the cluster. The final
center was chosen to minimize the square of the differences to these five galaxies.

To estimate the number of field galaxies that appear as interlopers in
the red-sequence, we performed an identical analysis using the
$i_{775}$ and $z_{850}$ images from the north and south HST GOODS
survey.  The GOODS images were processed using MultiDrizzle with the same
output pixel scale as the cluster fields.
The GOODS fields were observed in two different orientations
distinguished by a 45 degree relative rotation.
These observations roughly reproduce any PSF artifacts that may be induced
by the variety of telescope roll angles in the observations of the cluster fields.
Objects in the GOODS fields that pass the selection tests
described above were used to represent the population of interloper
field galaxies that would be expected to appear in any ACS pointing.

The red-sequence richness, $N_{red}$, was determined from the number
of red-sequence galaxies inside an aperture on the cluster center.  As
shown in Figure~\ref{fig:radprofile}, the surface density of interlopers
is approximately equal to the surface density of cluster galaxies 
at a distance $0.6$ Mpc from the cluster center.
More than $80\%$ of the galaxies in the
composite cluster are contained in a 0.6 Mpc radius aperture.
Essentially all red-sequence cluster members are included inside a
radius of 1.0 Mpc.  This aperture size is also fairly well matched to
the ACS field of view and was chosen to estimate the red-sequence
richnesses.

We sought a measure of the red-sequence richness that avoids
the Poisson noise from having very few members at the bright  
end of the luminosity function.
At the same time, we tried to provide a measure that represents the
clusters at all redshifts without substantial measurement error.  To
do so, we anchored the measurement at a magnitude $M\pr$ near the
faint end of the red sequence assuming linear passive evolution with
$dM/dz = -1$ in the rest frame luminosity.
$M\pr$ was chosen to be near the background dominated limit for the
most distant clusters in the sample, $M\pr = 24.5$ in the $z_{850}$
filter at $z=1.5$.  For comparison, in a model assuming passive
evolution, $z_{850} = 23.61$ for $L_*$ at $z=1.5$.
The red-sequence richness was then defined as the number of red-sequence
galaxies inside a 1 Mpc aperture with $M\pr - 2 < z_{850} < M\pr$.
Interloper field galaxies at other redshifts were removed by
subtracting the number of GOODS galaxies satisfying the same cut in
the same survey area.  Uncertainties in the background subtraction
were estimated from the Poisson statistics of the GOODS catalogs.
The uncertainty in the photometric redshifts of ISCS J1433.8+3325
and RCS033750-2844.8 could impact the measurements of $N_{red}$.
To test for additional systematic errors, $N_{red}$ was computed for
these two clusters at a redshift of $z = z_{phot}\pm 0.1$.
The measurements did not indicate any significant deviation from
the uncertainties estimated from the GOODS catalogs.